\documentclass[11pt,oneside]{article}
\usepackage{amssymb}
\newcommand{\text}[1]{\mbox{#1}}
\newcommand{\binom}[2]{{#1 \choose #2}}
\newcommand{\QATOP}[2]{{#1 \atop #2}}
\newcommand{\QATOPD}[2]{\left[ \begin{array}{c} #1 \\ #2
    \end{array} \right]}
\begin{document}

\author{Claudia Bauer\thanks{%
e-mail:Claudia.Bauer@physik.uni-muenchen.de} and Hartmut Wachter\thanks{%
e-mail:Hartmut.Wachter@physik.uni-muenchen.de} \\
Sektion Physik, Ludwig-Maximilians-Universit\"{a}t,\\
Theresienstr. 37, D-80333 M\"{u}nchen, Germany}
\title{Operator Representations on Quantum Spaces }
\maketitle

\begin{abstract}
In this article we present explicit formulae for q-differentiation on
quantum spaces which could be of particular importance in physics, i.e.,
q-deformed Minkowski space and q-deformed Euclidean space in three or four
dimensions. The calculations are based on the covariant differential
calculus of these quantum spaces. Furthermore, our formulae can be
regarded as a generalization of Jackson's q-derivative to three and
four dimensions.
\end{abstract}

\section{Introduction}

One might say the ideas of differential calculus are as old as physical
science itself. Since its invention by J. Newton and G.W. Leibniz there
hasn't been a necessity for an essential change. Although this can be seen
as a great success one cannot ignore the fact that up to now physicists
haven't been able to present a unified description of nature by using this
traditional tool, i.e., a theory which does not break down at any possible
space-time distances.

Quantum spaces, however, which are defined as co-module algebras of quantum
groups and which can be interpreted as deformations of ordinary co-ordinate
algebras \cite{RTF90} could provide a proper framework for developing a new
kind of non-commutative analysis \cite{Wes00}, \cite{Maj93}. For our \
purposes it is sufficient to consider a quantum space as an algebra $%
\mathcal{A}_{q}$ of formal power series in the non-commuting co-ordinates $%
X_{1},X_{2},\ldots ,X_{n}$

\begin{equation}
\mathcal{A}_{q}=\mathbb{C}\left[ \left[ X_{1},\ldots X_{n}\right] \right] /%
\mathcal{I}
\end{equation}
where $\mathcal{I}$ denotes the ideal generated by the relations of the
non-commuting co-ordinates.

The algebra $\mathcal{A}_{q}$ satisfies the Poincar\'{e}-Birkhoff-Witt
property, i.e., the dimension of the subspace of homogenous polynomials
should be the same as for commuting co-ordinates. This property is the
deeper reason why the monomials of normal ordering $X_{1}X_{2}\ldots X_{n}$
constitute a basis of $\mathcal{A}_{q}$. In particular, we can establish a
vector space isomorphism between $\mathcal{A}_{q}$ and the commutative
algebra $\mathcal{A}$ generated by ordinary co-ordinates $x_{1},x_{2},\ldots
,x_{n}$: 
\begin{eqnarray}
\mathcal{W} &:&\mathcal{A}\longrightarrow \mathcal{A}_{q}, \\
\mathcal{W}(x_{1}^{i_{1}}\ldots x_{n}^{i_{n}}) &=&X_{1}^{i_{1}}\ldots
X_{n}^{i_{n}}.  \nonumber
\end{eqnarray}
This vector space isomorphism can be extended to an algebra isomorphism
introducing a non-commutative product in $\mathcal{A}$, the so-called $\star 
$-product \cite{Moy49}, \cite{MSSW00}. This product is defined by the
relation 
\begin{equation}
\mathcal{W}(f\star g)=\mathcal{W}(f)\cdot \mathcal{W}(g)
\end{equation}
where $f$ and $g$ are formal power series in $\mathcal{A}$. In \cite{WW01}
we have calculated the $\star $-product for quantum spaces which could be of
particular importance in physics, i.e., q-deformed Minkowski space and
q-deformed Euclidean space in three or four dimensions.

Additionally, for each of these quantum spaces exists a symmetry algebra 
\cite{Dri85}, \cite{Jim85} and a covariant differential calculus \cite{WZ91}%
, which can provide an action upon the quantum spaces under consideration.
By means of the relation 
\begin{equation}
\mathcal{W}(h\triangleright f) \equiv h\triangleright \mathcal{W}(f)\text{,\quad }%
h\in \mathcal{H}\text{, }f\in \mathcal{A}\text{,}  \label{defrep}
\end{equation}
we are also able to introduce an action upon the corresponding commutative
algebra.

It is now our aim to present explicit formulae for the action of the partial
derivatives on these spaces. In addition we have worked out representations
of the generators of the q-deformed Lorentz algebra and the algebra of
q-deformed angular momentum in three or four dimensions. All explicit
formulae belong to left representations, as every right representation can
be deduced from a left one by applying some simple rules.

\section{q-Deformed Euclidean space in three dimensions\label{sec2}}

The q-deformed Euclidean space in three dimensions is spanned by the non-
commuting co-ordinates $X^{+},$ $X^{3},$ $X^{-}$. Their commutation
relations with the partial derivatives $\partial ^{+},$ $\partial ^{3},$ $%
\partial ^{-}$ can be written in the general form \cite{LWW97} 
\begin{equation}
\partial ^{A}X^{B}=g^{AB}+(\hat{R}^{-1})_{CD}^{AB}X^{C}\partial ^{D},\quad
A,B,C,D=3,\pm 
\end{equation}
where $\hat{R}^{-1}$ denotes the inverse of the R-matrix of the quantum
group SO$_{q}\left( 3\right)$ and $g^{AB}$ is the corresponding
metric. Explicitly we have 
\begin{eqnarray}
\partial ^{+}X^{+} &=&X^{+}\partial ^{+}, \\
\partial ^{+}X^{3} &=&q^{2}X^{3}\partial ^{+}-q^{2}\lambda \lambda
_{+}X^{+}\partial ^{3},  \nonumber \\
\partial ^{+}X^{-} &=&-q+q^{4}X^{-}\partial ^{+}-q^{3}\lambda \lambda
_{+}X^{3}\partial ^{3}+q^{3}\lambda ^{2}\lambda _{+}X^{+}\partial ^{-}, 
\nonumber \\[0.16in]
\partial ^{3}X^{+} &=&q^{2}X^{+}\partial ^{3}, \\
\partial ^{3}X^{3} &=&1+q^{2}X^{3}\partial ^{3}-q^{3}\lambda \lambda
_{+}X^{+}\partial ^{-},  \nonumber \\
\partial ^{3}X^{-} &=&q^{2}X^{-}\partial ^{3}-q^{2}\lambda \lambda
_{+}X^{3}\partial ^{-},  \nonumber \\[0.16in]
\partial ^{-}X^{+} &=&-q^{-1}+q^{4}X^{+}\partial ^{-}, \\
\partial ^{-}X^{3} &=&q^{2}X^{3}\partial ^{-},  \nonumber \\
\partial ^{-}X^{-} &=&X^{-}\partial ^{-}  \nonumber
\end{eqnarray}
with $\lambda =q-q^{-1}$ and $\lambda _{+}=q+q^{-1}$. On the q-deformed
version of 3-dimensional Euclidean space there exists a second covariant
differential calculus. Its defining relations read 
\begin{equation}
\hat{\partial}^{A}X^{B}=g^{AB}+(\hat{R})_{CD}^{AB}X^{C}\hat{\partial}%
^{D},\quad A,B,C,D=3,\pm .
\end{equation}
Written out, we get 
\begin{eqnarray}
\hat{\partial}^{+}X^{+} &=&X^{+}\hat{\partial}^{+}, \\
\hat{\partial}^{+}X^{3} &=&q^{-2}X^{3}\hat{\partial}^{+},  \nonumber \\
\hat{\partial}^{+}X^{-} &=&-q+q^{-4}X^{-}\hat{\partial}^{+},  \nonumber \\%
[0.16in]
\hat{\partial}^{3}X^{+} &=&q^{-2}X^{+}\hat{\partial}^{3}+q^{-2}\lambda
\lambda _{+}X^{3}\hat{\partial}^{+}, \\
\hat{\partial}^{3}X^{3} &=&1+q^{-2}X^{3}\hat{\partial}^{3}+q^{-3}\lambda
\lambda _{+}X^{-}\hat{\partial}^{+},  \nonumber \\
\hat{\partial}^{3}X^{-} &=&q^{-2}X^{-}\hat{\partial}^{3},  \nonumber \\%
[0.16in]
\hat{\partial}^{-}X^{+} &=&-q^{-1}+q^{-4}X^{+}\hat{\partial}%
^{-}+q^{-3}\lambda \lambda _{+}X^{3}\hat{\partial}^{3}+q^{-3}\lambda
^{-2}\lambda _{+}X^{-}\hat{\partial}^{+}, \\
\hat{\partial}^{-}X^{3} &=&q^{-2}X^{3}\hat{\partial}^{-}+q^{-2}\lambda
\lambda _{+}X^{-}\hat{\partial}^{3},  \nonumber \\
\hat{\partial}^{-}X^{-} &=&X^{-}\hat{\partial}^{-}.  \nonumber
\end{eqnarray}

These relations yield a Hopf structure for the two sets of derivatives which
can be derived by the same method already explained in \cite{OSWZ92}. Using
the generators of angular momentum\footnote{%
One has to keep attention on the different normalisation of the $L^{A}$.
Analogous to \cite{Blo01} we made the substitution $L^{A}\rightarrow
-q^{-3}L^{A}.$} $L^{+},$ $L^{-},$ $\tau ^{-1/2}$ and the scaling operator $%
\Lambda $ \cite{LWW97} one obtains in the first case for the co-product $%
\Delta ,$ the antipode $S$ and the co-unit $\varepsilon $ the following
expressions: 
\begin{eqnarray}
\Delta (\partial ^{-}) &=&\partial ^{-}\otimes 1+\Lambda ^{\frac{1}{2}}\tau
^{-\frac{1}{2}}\otimes \partial ^{-}, \\
\Delta (\partial ^{3}) &=&\partial ^{3}\otimes 1+\Lambda ^{\frac{1}{2}%
}\otimes \partial ^{3}+\lambda \lambda _{+}\Lambda ^{\frac{1}{2}%
}L^{+}\otimes \partial ^{-},  \nonumber \\
\Delta (\partial ^{+}) &=&\partial ^{+}\otimes 1+\Lambda ^{\frac{1}{2}}\tau
^{\frac{1}{2}}\otimes \partial ^{+}+q\lambda \lambda _{+}\Lambda ^{\frac{1}{2%
}}\tau ^{\frac{1}{2}}L^{+}\otimes \partial ^{3}  \nonumber \\
&&+\,q^{2}\lambda ^{2}\lambda _{+}\Lambda ^{\frac{1}{2}}\tau ^{\frac{1}{2}%
}(L^{+})^{2}\otimes \partial ^{-},  \nonumber \\[0.16in]
S(\partial ^{-}) &=&-\Lambda ^{-\frac{1}{2}}\tau ^{\frac{1}{2}}\partial ^{-},
\\
S(\partial ^{3}) &=&-\Lambda ^{-\frac{1}{2}}\partial ^{3}+q^{2}\lambda
\lambda _{+}\Lambda ^{-\frac{1}{2}}\tau ^{\frac{1}{2}}L^{+}\partial ^{-}, 
\nonumber \\
S(\partial ^{+}) &=&-\Lambda ^{-\frac{1}{2}}\tau ^{-\frac{1}{2}}\partial
^{+}+q\lambda \lambda _{+}\Lambda ^{-\frac{1}{2}}L^{+}\partial
^{3}-q^{4}\lambda ^{2}\lambda _{+}\Lambda ^{-\frac{1}{2}}\tau ^{\frac{1}{2}%
}(L^{+})^{2}\partial ^{-},  \nonumber \\[0.16in]
\varepsilon (\partial ^{+}) &=&\varepsilon (\partial ^{3})=\varepsilon
(\partial ^{-})=0.
\end{eqnarray} 
In the second case the Hopf structure is given by 
\begin{eqnarray}
\Delta (\hat{\partial}^{+}) &=&\hat{\partial}^{+}\otimes 1+\Lambda ^{-\frac{1%
}{2}}\tau ^{-\frac{1}{2}}\otimes \hat{\partial}^{+}, \\
\Delta (\hat{\partial}^{3}) &=&\hat{\partial}^{3}\otimes 1+\Lambda ^{-\frac{1%
}{2}}\otimes \hat{\partial}^{3}+\lambda \lambda _{+}\Lambda ^{-\frac{1}{2}%
}L^{-}\otimes \hat{\partial}^{+},  \nonumber \\
\Delta (\hat{\partial}^{-}) &=&\hat{\partial}^{-}\otimes 1+\Lambda ^{-\frac{1%
}{2}}\tau ^{-\frac{1}{2}}\otimes \hat{\partial}^{-}+q^{-1}\lambda \lambda
_{+}\Lambda ^{-\frac{1}{2}}\tau ^{\frac{1}{2}}L^{-}\otimes \hat{\partial}^{3}
\nonumber \\
&&+\,q^{-2}\lambda ^{2}\lambda _{+}\Lambda ^{-\frac{1}{2}}\tau ^{\frac{1}{2}%
}(L^{-})^{2}\otimes \hat{\partial}^{+},  \nonumber \\[0.16in]
S(\hat{\partial}^{+}) &=&-\Lambda ^{\frac{1}{2}}\tau ^{\frac{1}{2}}\hat{%
\partial}^{+}, \\
S(\hat{\partial}^{3}) &=&-\Lambda ^{\frac{1}{2}}\hat{\partial}%
^{3}+q^{-2}\lambda \lambda _{+}\Lambda ^{\frac{1}{2}}\tau ^{\frac{1}{2}}L^{-}%
\hat{\partial}^{+},  \nonumber \\
S(\hat{\partial}^{-}) &=&-\Lambda ^{\frac{1}{2}}\tau ^{-\frac{1}{2}}\hat{%
\partial}^{-}+q^{-1}\lambda \lambda _{+}\Lambda ^{\frac{1}{2}}L^{-}\hat{%
\partial}^{3}-q^{-4}\lambda ^{2}\lambda _{+}\Lambda ^{\frac{1}{2}}\tau ^{%
\frac{1}{2}}(L^{-})^{2}\hat{\partial}^{+},  \nonumber \\[0.16in]
\varepsilon (\hat{\partial}^{+}) &=&\varepsilon (\hat{\partial}%
^{3})=\varepsilon (\hat{\partial}^{-})=0.
\end{eqnarray}
Due to the relation 
\begin{equation}
\partial ^{A}\triangleright (f\star g)=(\partial _{\left( 1\right)
}^{A}\triangleright f)\star (\partial _{\left( 2\right) }^{A}\triangleright
g),
\end{equation}
the Leibniz rules for products of arbitrary power series can be read off
from the co-product $\Delta (\partial ^{A})=\partial _{\left( 1\right)
}^{A}\otimes \partial _{\left( 2\right) }^{A}$ quite easily \cite{KS97}, 
\cite{Maj95}.

For applying this formula, however, it is necessary to know the
representations of the generators $L^{+},$ $L^{-},$ $\tau ^{-1/2}$ and the
scaling operator $\Lambda ,$ which can be computed from the commutation
relations 
\begin{eqnarray}
L^{+}X^{+} &=&X^{+}L^{+}, \\
L^{+}X^{3} &=&X^{3}L^{+}-qX^{+}\tau ^{-\frac{1}{2}},  \nonumber \\
L^{+}X^{-} &=&X^{-}L^{+}-X^{3}\tau ^{-\frac{1}{2}},  \nonumber \\[0.16in]
L^{-}X^{+} &=&X^{+}L^{-}+X^{3}\tau ^{-\frac{1}{2}}, \\
L^{-}X^{3} &=&X^{3}L^{-}+q^{-1}X^{-}\tau ^{-\frac{1}{2}},  \nonumber \\
L^{-}X^{-} &=&X^{-}L^{-},  \nonumber \\[0.16in]
\tau ^{-\frac{1}{2}}X^{\pm } &=&q^{\pm 2}X^{\pm }\tau ^{-\frac{1}{2}},\\
\tau ^{-\frac{1}{2}}X^{3} &=&X^{3}\tau ^{-\frac{1}{2}},\nonumber \\
\Lambda ^{\frac{1}{2}}X^{A} &=&q^{2}X^{A}\Lambda ^{\frac{1}{2}},
\quad A=\pm ,3.
\end{eqnarray}
To calculate the explicit form of their action on the Quantum space algebra
we iterate the action of the generators on monomials of normal ordering $%
X^{+}X^{3}X^{-}$ until all generators have
moved to the right. With the relation $T\triangleright 1=\varepsilon (T)$
and after a possible normal ordering the wanted representations follow
immediately. Such calculations can also be found in \cite{Schra00}. Finally,
in the sense of definition (\ref{defrep}) the action of the generators $%
L^{+},$ $L^{-},$ $\tau ^{-1/2}$ \ and the scaling operator $\Lambda $ take
the form\footnote{%
For notation see appendix \ref{AppA}.} 
\begin{eqnarray}
L^{+}\triangleright f
&=&-q^{2}x^{3}(D_{q^{4}}^{-}f)(q^{-2}x^{-})-qx^{+}(D_{q^{2}}^{3}f)(q^{-2}x^{-}),
\\
L^{-}\triangleright f
&=&x^{3}(D_{q^{4}}^{+}f)(q^{-2}x^{-})+q^{-1}x^{-}(D_{q^{2}}^{3}f)(q^{-2}x^{-}),
\nonumber \\
\tau ^{\pm \frac{1}{2}}\triangleright f &=&f(q^{\mp 2}x^{+},q^{\pm 2}x^{-}),
\nonumber \\
\Lambda ^{\pm \frac{1}{2}}\triangleright f &=&f(q^{\pm 2}x^{+},q^{\pm
2}x^{3},q^{\pm 2}x^{-}).  \nonumber
\end{eqnarray}
Similar expressions can be derived for the partial derivatives $\partial
^{+},$ $\partial ^{3},$ $\partial ^{-}$ with the end result 
\begin{eqnarray}
\partial ^{-}\triangleright f &=&-q^{-1}D_{q^{4}}^{+}f, \\
\partial ^{3}\triangleright f &=&D_{q^{2}}^{3}f(q^{2}x^{+}),  \nonumber \\
\partial ^{+}\triangleright f &=&-qD_{q^{4}}^{-}f(q^{2}x^{3})-q\lambda
x^{+}(D_{q^{2}}^{3})^{2}f.  \nonumber
\end{eqnarray}
In the case of the second differential calculus the representations of the
partial derivatives take on a very simple form, if they refer to the
ordering $X^{-}X^{3}X^{+}$. For this new
ordering we have 
\begin{eqnarray}
\hat{\partial}^{+}\tilde{\triangleright}f &=&-qD_{q^{-4}}^{-}f, \\
\hat{\partial}^{3}\tilde{\triangleright}f &=&D_{q^{-2}}^{3}f(q^{-2}x^{-}), 
\nonumber \\
\hat{\partial}^{-}\tilde{\triangleright}f
&=&-q^{-1}D_{q^{-4}}^{+}f(q^{-2}x^{3})+q^{-1}\lambda
x^{-}(D_{q^{-2}}^{3})^{2}f.  \nonumber
\end{eqnarray}
In addition, we give the identities 
\begin{eqnarray}
\partial ^{A}\triangleright (\hat{U}^{-1}f) &=&\hat{U}^{-1}(\partial ^{A}%
\tilde{\triangleright}f), \\
\hat{\partial}^{A}\tilde{\triangleright}(\hat{U}f) &=&\hat{U}(\hat{\partial}%
^{A}\triangleright f)  \nonumber
\end{eqnarray}
where
\footnote{%
For notation see again appendix\ \ref{AppA}} 
\begin{eqnarray}
\hat{U}^{-1}f  
&=&\sum_{i=0}^{\infty }\lambda ^{i}\frac{(x^{3})^{2i}}{[[i]]_{q^{4}}!}q^{2%
\hat{n}_{3}\left( \hat{n}_{+}+\hat{n}_{-}+i\right) }\left(
D_{q^{4}}^{+}D_{q^{4}}^{-}\right) ^{i}f,\label{Um3dimIn} \\[0.16in]
\hat{U}f 
&=&\sum_{i=0}^{\infty }\left( -\lambda \right) ^{i}\frac{(x^{3})^{2i}}{%
[[i]]_{q^{-4}}!}q^{-2\hat{n}_{3}\left( \hat{n}_{+}+\hat{n}_{-}+i\right)
}\left( D_{q^{-4}}^{+}D_{q^{-4}}^{-}\right) ^{i}f.\label{Um3dim} 
\end{eqnarray}
With these formulae at hand which can easily be derived from the
considerations in \cite{WW01} we are in a position to deal with representations
of one given ordering only, as the operators $\hat{U}^{-1}$ and $\hat{U}$
transform functions of ordering $%
X^{-}X^{3}X^{+}$ to the corresponding ones of
ordering $X^{+}X^{3}X^{-}$ and vice versa.

All representations considered so far have been computed by commuting the
acting generators from the left side of a monomial to the right. These
representations are thus called left representations. However, if we commute
the acting generators from the right side of a monomial to the left, right
representations will consequently arise. But these right representations can
be read off from left ones quite easily, as right representations 
are always linked to left ones via the
identity 
\begin{equation}
\overline{\partial ^{A}\triangleright f}=\overline{f}\triangleleft \overline{%
\partial ^{A}}.
\end{equation}
From the conjugation properties \cite{LWW97} 
\begin{equation}
\overline{X^{+}}=-qX^{-},\quad \overline{X^{3}}=X^{3},\quad \overline{X^{-}}%
=-q^{-1}X^{+},
\end{equation} 
\[
\overline{\partial ^{+}}=q^{-5}\hat{\partial}^{-},\quad \overline{\partial
^{3}}=-q^{-6}\hat{\partial}^{3},\quad \overline{\partial ^{-}}=q^{-7}\hat{%
\partial}^{+},
\]
\[
\overline{L^{+}}=-qL^{-},\quad \overline{L^{-}}=-q^{-1}L^{+}
\]
one obtains the translation rules 
\begin{eqnarray}
f\triangleleft L^{+}&\stackrel{+\leftrightarrow -}{\longleftrightarrow }&%
L^{-}\triangleright f, \\
f\triangleleft L^{-}&\stackrel{+\leftrightarrow -}{\longleftrightarrow }&%
L^{+}\triangleright f,  \nonumber \\[0.16in]
f\triangleleft \partial ^{+}&\stackrel{+\leftrightarrow -}{%
\longleftrightarrow }&-q^{-6}\hat{\partial}^{-}\triangleright f, \\
f\triangleleft \partial ^{-}&\stackrel{+\leftrightarrow -}{%
\longleftrightarrow }&-q^{-6}\hat{\partial}^{+}\triangleright f,  \nonumber \\
f\triangleleft \partial ^{3}&\stackrel{+\leftrightarrow -}{%
\longleftrightarrow }&-q^{-6}\hat{\partial}^{3}\triangleright f,  \nonumber \\%
[0.16in]
f\triangleleft \hat{\partial}^{+}&\stackrel{+\leftrightarrow -}{%
\longleftrightarrow }&-q^{6}\partial ^{-}\triangleright f, \\
f\triangleleft \hat{\partial}^{-}&\stackrel{+\leftrightarrow -}{%
\longleftrightarrow }&-q^{6}\partial ^{+}\triangleright f,  \nonumber \\
f\triangleleft \hat{\partial}^{3}&\stackrel{+\leftrightarrow -}{%
\longleftrightarrow }&-q^{6}\partial ^{3}\triangleright f  \nonumber
\end{eqnarray}
where the symbol $\stackrel{+\leftrightarrow -}{\longleftrightarrow }$
denotes that one can make a transition between the two expressions by
applying the substitutions 
\begin{equation}
x^{\pm }\rightarrow x^{\mp },\quad D_{q^{a}}^{\pm }\rightarrow
D_{q^{a}}^{\mp }\quad \hat{n}^{\pm }\rightarrow \hat{n}^\mp .
\end{equation}
The following shall serve as an example: 
\[
x^{+}x^{-}(D_{q}^{+})^{2}D_{q}^{-}f(q^{2}x^{-})\stackrel{+\leftrightarrow -}{%
\longleftrightarrow }x^{-}x^{+}(D_{q}^{-})^{2}D_{q}^{+}f(q^{2}x^{+}).
\]
The right representations of the generators $\tau ^{3}$ and $\Lambda $ are
derived most easily from the identity 
\begin{equation}
f\triangleleft h=S^{-1}(h)\triangleright f,  \label{linksdar}
\end{equation}
hence 
\begin{eqnarray}
f\triangleleft \tau ^{\pm \frac{1}{2}} &=&S^{-1}(\tau ^{\pm \frac{1}{2}%
})\triangleright f=\tau ^{\pm \frac{1}{2}}\triangleright f, \\
f\triangleleft \Lambda ^{\pm \frac{1}{2}} &=&S^{-1}(\Lambda ^{\pm \frac{1}{2}%
})\triangleright f=\Lambda ^{\mp \frac{1}{2}}\triangleright f.  \nonumber
\end{eqnarray}
Finally, let us remark that due to the relation 
\begin{equation}
f\partial ^{A}=\partial _{\left( 2\right) }^{A}(f\triangleleft \partial
_{\left( 1\right) }^{A})
\end{equation}
again the co-products of the two differential calculi directly yield 
Leibniz rules for right representations of the partial derivatives.

\section{q-Deformed Euclidean space in four dimensions}

The 4-dimensional q-deformed Euclidean space can be treated in very much the
same way as the 3-dimensional case. For the relations between partial
derivatives and coordinates we now have 
\begin{equation}
\partial ^{i}X^{j}=g^{ij}+q(\hat{R}^{-1})_{kl}^{ij}X^{k}\partial ^{l},\quad
i,j,k,l=1,\ldots ,4
\end{equation}
where $g^{ij}$ denotes the 4-dimensional Euclidean Quantum space metric and $%
\hat{R}$ the R-matrix of $SO_{q}(4).$ With the notation in \cite{KS97} these
relations read explicitly 
\begin{eqnarray}
\partial ^{1}X^{1} &=&X^{1}\partial ^{1}, \\
\partial ^{1}X^{2} &=&qX^{2}\partial ^{1},  \nonumber \\
\partial ^{1}X^{3} &=&qX^{3}\partial ^{1},  \nonumber \\
\partial ^{1}X^{4} &=&q^{-1}+q^{2}X^{4}\partial ^{1},  \nonumber \\[0.16in]
\partial ^{2}X^{1} &=&qX^{1}\partial ^{2}-q\lambda X^{2}\partial ^{1}, \\
\partial ^{2}X^{2} &=&X^{2}\partial ^{2},  \nonumber \\
\partial ^{2}X^{3} &=&1+q^{2}X^{3}\partial ^{2}+q^{2}\lambda X^{4}\partial
^{1},  \nonumber \\
\partial ^{2}X^{4} &=&qX^{4}\partial ^{2},  \nonumber \\[0.16in]
\partial ^{3}X^{1} &=&qX^{1}\partial ^{3}-q\lambda X^{3}\partial ^{1}, \\
\partial ^{3}X^{2} &=&1+q^{2}X^{2}\partial ^{3}+q^{2}\lambda X^{4}\partial
^{1},  \nonumber \\
\partial ^{3}X^{3} &=&X^{3}\partial ^{3},  \nonumber \\
\partial ^{3}X^{4} &=&qX^{4}\partial ^{3},  \nonumber \\[0.16in]
\partial ^{4}X^{1} &=&q+q^{2}X^{1}\partial ^{4}+q^{2}\lambda \left(
X^{2}\partial ^{3}+X^{3}\partial ^{2}+\lambda X^{4}\partial ^{1}\right) , \\
\partial ^{4}X^{2} &=&qX^{2}\partial ^{4}-q\lambda X^{4}\partial ^{2}, 
\nonumber \\
\partial ^{4}X^{3} &=&qX^{3}\partial ^{4}-q\lambda X^{4}\partial ^{3}, 
\nonumber \\
\partial ^{4}X^{4} &=&X^{4}\partial ^{4}.  \nonumber
\end{eqnarray}
For the second set of derivatives the following relations hold: 
\begin{equation}
\partial ^{i}X^{j}=g^{ij}+q^{-1}(\hat{R})_{kl}^{ij}X^{k}\partial ^{l},\quad
i,j,k,l=1,\ldots ,4.
\end{equation}
In a more explicit form one can write 
\begin{eqnarray}
\hat{\partial}^{1}X^{1} &=&X^{1}\hat{\partial}^{1}, \\
\hat{\partial}^{1}X^{2} &=&q^{-1}X^{2}\hat{\partial}^{1}+q^{-1}\lambda X^{1}%
\hat{\partial}^{2},  \nonumber \\
\hat{\partial}^{1}X^{3} &=&q^{-1}X^{3}\hat{\partial}^{1}+q^{-1}\lambda X^{1}%
\hat{\partial}^{3},  \nonumber \\
\hat{\partial}^{1}X^{4} &=&q^{-1}+q^{-2}X^{4}\hat{\partial}%
^{1}-q^{-2}\lambda (X^{2}\hat{\partial}^{3}+X^{3}\hat{\partial}^{2}-\lambda
X^{1}\hat{\partial}^{4}),  \nonumber \\[0.16in]
\hat{\partial}^{2}X^{1} &=&q^{-1}X^{1}\hat{\partial}^{2}, \\
\hat{\partial}^{2}X^{2} &=&X^{2}\hat{\partial}^{2},  \nonumber \\
\hat{\partial}^{2}X^{3} &=&1+q^{-2}X^{3}\hat{\partial}^{2}-q^{-2}\lambda
X^{1}\hat{\partial}^{4},  \nonumber \\
\hat{\partial}^{2}X^{4} &=&q^{-1}X^{4}\hat{\partial}^{2}+q^{-1}\lambda X^{2}%
\hat{\partial}^{4},  \nonumber \\[0.16in]
\hat{\partial}^{3}X^{1} &=&q^{-1}X^{1}\hat{\partial}^{3}, \\
\hat{\partial}^{3}X^{2} &=&1+q^{-2}X^{2}\hat{\partial}^{3}-q^{-2}\lambda
X^{1}\hat{\partial}^{4},  \nonumber \\
\hat{\partial}^{3}X^{3} &=&X^{3}\hat{\partial}^{3},  \nonumber \\
\hat{\partial}^{3}X^{4} &=&q^{-1}X^{4}\hat{\partial}^{3}+q^{-1}\lambda X^{3}%
\hat{\partial}^{4},  \nonumber \\[0.16in]
\hat{\partial}^{4}X^{1} &=&q+q^{-2}X^{1}\hat{\partial}^{4}, \\
\hat{\partial}^{4}X^{2} &=&q^{-1}X^{2}\hat{\partial}^{4},  \nonumber \\
\hat{\partial}^{4}X^{3} &=&q^{-1}X^{3}\hat{\partial}^{4},  \nonumber \\
\hat{\partial}^{4}X^{4} &=&X^{4}\hat{\partial}^{4}.  \nonumber
\end{eqnarray}
From these relations we again can deduce a Hopf structure for the
derivatives $\partial ^{i},$ $i=1,\ldots ,4,$ which in terms of the $%
U_{q}( so_4) $ generators $L_{i}^{\pm },$ $K_{i}$ $(i=1,2)$ and the
scaling operator $\Lambda $ becomes 
\begin{eqnarray}
\Delta (\partial ^{1}) &=&\partial ^{1}\otimes 1+\Lambda ^{\frac{1}{2}%
}K_{1}^{\frac{1}{2}}K_{2}^{\frac{1}{2}}\otimes \partial ^{1}, \\
\Delta (\partial ^{2}) &=&\partial ^{2}\otimes 1+\Lambda ^{\frac{1}{2}%
}K_{1}^{-\frac{1}{2}}K_{2}^{\frac{1}{2}}\otimes \partial ^{2}  \nonumber \\
&&+\,q\lambda \Lambda ^{\frac{1}{2}}K_{1}^{\frac{1}{2}}K_{2}^{\frac{1}{2}%
}L_{1}^{+}\otimes \partial ^{1},  \nonumber \\
\Delta (\partial ^{3}) &=&\partial ^{3}\otimes 1+\Lambda ^{\frac{1}{2}%
}K_{1}^{\frac{1}{2}}K_{2}^{-\frac{1}{2}}\otimes \partial ^{3}  \nonumber \\
&&+\,q\lambda \Lambda ^{\frac{1}{2}}K_{1}^{\frac{1}{2}}K_{2}^{\frac{1}{2}%
}L_{2}^{+}\otimes \partial ^{1},  \nonumber \\
\Delta (\partial ^{4}) &=&\partial ^{4}\otimes 1+\Lambda ^{\frac{1}{2}%
}K_{1}^{-\frac{1}{2}}K_{2}^{-\frac{1}{2}}\otimes \partial ^{4}  \nonumber \\
&&-\,q^{2}\lambda ^{2}\Lambda ^{\frac{1}{2}}K_{1}^{\frac{1}{2}}K_{2}^{\frac{1%
}{2}}L_{1}^{+}L_{2}^{+}\otimes \partial ^{1}  \nonumber \\
&&-\,q\lambda \Lambda ^{\frac{1}{2}}K_{1}^{-\frac{1}{2}}K_{2}^{\frac{1}{2}%
}L_{2}^{+}\otimes \partial ^{2}  \nonumber \\
&&-\,q\lambda \Lambda ^{\frac{1}{2}}K_{1}^{\frac{1}{2}}K_{2}^{-\frac{1}{2}%
}L_{1}^{+}\otimes \partial ^{3},  \nonumber \\[0.16in]
S(\partial ^{1}) &=&-\Lambda ^{-\frac{1}{2}}K_{1}^{-\frac{1}{2}}K_{2}^{-%
\frac{1}{2}}\partial ^{1}, \\
S(\partial ^{2}) &=&-\Lambda ^{-\frac{1}{2}}K_{1}^{\frac{1}{2}}K_{2}^{-\frac{%
1}{2}}(\partial ^{2}-q^{2}\lambda L_{1}^{+}\partial ^{1}),  \nonumber \\
S(\partial ^{3}) &=&-\Lambda ^{-\frac{1}{2}}K_{1}^{-\frac{1}{2}}K_{2}^{\frac{%
1}{2}}(\partial ^{3}-q^{2}\lambda L_{2}^{+}\partial ^{1}),  \nonumber \\
S(\partial ^{4}) &=&-\Lambda ^{-\frac{1}{2}}K_{1}^{\frac{1}{2}}K_{2}^{\frac{1%
}{2}}(\partial ^{4}+q^{2}\lambda (L_{1}^{+}\partial ^{3}+L_{2}^{+}\partial
^{2}))  \nonumber \\
&&-\,q^{4}\lambda ^{2}\Lambda ^{-\frac{1}{2}}K_{1}^{\frac{1}{2}}K_{2}^{\frac{%
1}{2}}L_{1}^{+}L_{2}^{+}\partial ^{1},  \nonumber \\[0.16in]
\varepsilon (\partial ^{1}) &=&\varepsilon (\partial ^{2})=\varepsilon
(\partial ^{3})=\varepsilon (\partial ^{4})=0.
\end{eqnarray}
And in the same manner we get for the other derivatives $\hat{\partial}^{i},$ 
$i=1,\ldots ,4$%
\begin{eqnarray}
\Delta (\hat{\partial}^{1}) &=&\hat{\partial}^{1}\otimes 1+\Lambda ^{-\frac{1%
}{2}}K_{1}^{-\frac{1}{2}}K_{2}^{-\frac{1}{2}}\otimes \hat{\partial}^{1} \\
&&-\,q^{-2}\lambda ^{2}\Lambda ^{-\frac{1}{2}}K_{1}^{\frac{1}{2}}K_{2}^{%
\frac{1}{2}}L_{1}^{-}L_{2}^{-}\otimes \hat{\partial}^{4}  \nonumber \\
&&-\,q^{-1}\lambda \Lambda ^{-\frac{1}{2}}K_{1}^{\frac{1}{2}}K_{2}^{-\frac{1%
}{2}}L_{1}^{-}\otimes \hat{\partial}^{2}  \nonumber \\
&&-\,q^{-1}\lambda \Lambda ^{-\frac{1}{2}}K_{1}^{-\frac{1}{2}}K_{2}^{\frac{1%
}{2}}L_{2}^{-}\otimes \hat{\partial}^{3},  \nonumber \\
\Delta (\hat{\partial}^{2})) &=&\hat{\partial}^{2}\otimes 1+\Lambda ^{-\frac{%
1}{2}}K_{1}^{\frac{1}{2}}K_{2}^{-\frac{1}{2}}\otimes \hat{\partial}^{2} \\
&&+\,q^{-1}\lambda \Lambda ^{-\frac{1}{2}}K_{1}^{\frac{1}{2}}K_{2}^{\frac{1}{%
2}}L_{2}^{-}\otimes \hat{\partial}^{4},  \nonumber \\
\Delta (\hat{\partial}^{3}) &=&\hat{\partial}^{3}\otimes 1+\Lambda ^{-\frac{1%
}{2}}K_{1}^{-\frac{1}{2}}K_{2}^{\frac{1}{2}}\otimes \hat{\partial}^{3} 
\nonumber \\
&&+\,q^{-1}\lambda \Lambda ^{-\frac{1}{2}}K_{1}^{\frac{1}{2}}K_{2}^{\frac{1}{%
2}}L_{1}^{-}\otimes \hat{\partial}^{4},  \nonumber \\
\Delta (\hat{\partial}^{4}) &=&\hat{\partial}^{4}\otimes 1+\Lambda ^{-\frac{1%
}{2}}K_{1}^{\frac{1}{2}}K_{2}^{\frac{1}{2}}\otimes \hat{\partial}^{4}, 
\nonumber \\[0.16in]
S(\hat{\partial}^{1}) &=&-\Lambda ^{\frac{1}{2}}K_{1}^{\frac{1}{2}}K_{2}^{%
\frac{1}{2}}(\hat{\partial}^{1}+q^{-2}\lambda (L_{1}^{-}\hat{\partial}%
^{2}+L_{2}^{-}\hat{\partial}^{3})) \\
&&+\,q^{-4}\lambda ^{2}\Lambda ^{\frac{1}{2}}K_{1}^{\frac{1}{2}}K_{2}^{\frac{%
1}{2}}L_{1}^{-}L_{2}^{-}\hat{\partial}^{4},  \nonumber \\
S(\hat{\partial}^{2}) &=&-\Lambda ^{\frac{1}{2}}K_{1}^{-\frac{1}{2}}K_{2}^{%
\frac{1}{2}}(\hat{\partial}^{2}-q^{-2}\lambda L_{2}^{-}\hat{\partial}^{4}), 
\nonumber \\
S(\hat{\partial}^{3}) &=&-\Lambda ^{\frac{1}{2}}K_{1}^{-\frac{1}{2}}K_{2}^{-%
\frac{1}{2}}(\hat{\partial}^{3}-q^{-2}\lambda L_{1}^{-}\hat{\partial}^{4}), 
\nonumber \\
S(\hat{\partial}^{4}) &=&-\Lambda ^{\frac{1}{2}}K_{1}^{-\frac{1}{2}}K_{2}^{-%
\frac{1}{2}}\hat{\partial}^{4},  \nonumber \\[0.16in]
\varepsilon (\hat{\partial}^{1}) &=&\varepsilon (\hat{\partial}%
^{2})=\varepsilon (\hat{\partial}^{3})=\varepsilon (\hat{\partial}^{4})=0.
\end{eqnarray}

Note that the above expressions again yield Leibniz rules for products of
functions, if the representations of the given $U_q(so_{4}) $
generators and the $\Lambda $-operator are known. Toward this end we need
their commutation relations with the Quantum space coordinates, for which we
have \cite{Oca96} 
\begin{eqnarray}
L_{1}^{+}X^{1} &=&qX^{1}L_{1}^{+}-q^{-1}X^{2}, \\
L_{1}^{+}X^{2} &=&q^{-1}X^{2}L_{1}^{+},  \nonumber \\
L_{1}^{+}X^{3} &=&qX^{3}L_{1}^{+}+q^{-1}X^{4},  \nonumber \\
L_{1}^{+}X^{4} &=&q^{-1}X^{4}L_{1}^{+},  \nonumber \\[0.16in]
L_{2}^{+}X^{1} &=&qX^{1}L_{2}^{+}-q^{-1}X^{3}, \\
L_{2}^{+}X^{2} &=&qX^{2}L_{2}^{+}+q^{-1}X^{4},  \nonumber \\
L_{2}^{+}X^{3} &=&q^{-1}X^{3}L_{2}^{+},  \nonumber \\
L_{2}^{+}X^{4} &=&q^{-1}X^{4}L_{2}^{+},  \nonumber \\[0.16in]
L_{1}^{-}X^{1} &=&qX^{1}L_{1}^{-}, \\
L_{1}^{-}X^{2} &=&q^{-1}X^{2}L_{1}^{-}-qX^{1},  \nonumber \\
L_{1}^{-}X^{3} &=&qX^{3}L_{1}^{-},  \nonumber \\
L_{1}^{-}X^{4} &=&q^{-1}X^{4}L_{1}^{-}+qX^{3},  \nonumber \\[0.16in]
L_{2}^{-}X^{1} &=&qX^{1}L_{2}^{-}, \\
L_{2}^{-}X^{2} &=&qX^{2}L_{2}^{-},  \nonumber \\
L_{2}^{-}X^{3} &=&q^{-1}X^{3}L_{2}^{-}-qX^{1},  \nonumber \\
L_{2}^{-}X^{4} &=&q^{-1}X^{4}L_{2}^{-}+qX^{2},  \nonumber \\[0.16in]
K_{1}X^{1} &=&q^{-1}X^{1}K_{1}, \\
K_{1}X^{2} &=&qX^{2}K_{1},  \nonumber \\
K_{1}X^{3} &=&q^{-1}X^{3}K_{1},  \nonumber \\
K_{1}X^{4} &=&qX^{4}K_{1},  \nonumber \\[0.16in]
K_{2}X^{1} &=&q^{-1}X^{1}K_{2}, \\
K_{2}X^{2} &=&q^{-1}X^{2}K_{2},  \nonumber \\
K_{2}X^{3} &=&qX^{3}K_{2},  \nonumber \\
K_{2}X^{4} &=&qX^{4}K_{2},  \nonumber \\[0.16in]
\Lambda X^{i} &=&q^{2}X^{i}\Lambda ,\quad i=1,\ldots ,4.
\end{eqnarray}
In normal ordering $X^{1}X^{2}X^{3}X^{4}$ these
relations lead to left representations of the following form: 
\begin{eqnarray}
L_{1}^{+}\triangleright f
&=&x^{4}D_{q^{2}}^{3}f(qx^{1},q^{-1}x^{2},q^{-1}x^{3})-x^{2}D_{q^{2}}^{1}f(q^{-1}x^{1}),
\\
L_{2}^{+}\triangleright f
&=&x^{4}D_{q^{2}}^{2}f(qx^{1},q^{-1}x^{2},q^{-1}x^{3})-x^{3}D_{q^{2}}^{1}f(q^{-1}x^{1}),
\nonumber \\
L_{1}^{-}\triangleright f
&=&qx^{3}D_{q^{-2}}^{4}f(qx^{1},q^{-1}x^{2},qx^{3})-qx^{1}D_{q^{-2}}^{2}f(qx^{1}),
\nonumber \\
L_{2}^{-}\triangleright f
&=&qx^{2}D_{q^{-2}}^{4}f(qx^{1},qx^{2},q^{-1}x^{3})-qx^{1}D_{q^{-2}}^{3}f(qx^{1}),
\nonumber \\[0.16in]
K_{1}\triangleright f &=&f(q^{-1}x^{1},qx^{2},q^{-1}x^{3},qx^{4}), \\
K_{2}\triangleright f &=&f(q^{-1}x^{1},q^{-1}x^{2},qx^{3},qx^{4}),  \nonumber
\\
\Lambda ^{\pm \frac{1}{2}}\triangleright f &=&f(q^{\pm 1}x^{1},q^{\pm
1}x^{2},q^{\pm 1}x^{3},q^{\pm 1}x^{4}).  \nonumber
\end{eqnarray}
Accordingly, the representations of the partial derivatives $\hat{\partial}%
^{i}$ can be written as 
\begin{eqnarray}
\hat{\partial}^{1}\triangleright f
&=&q^{-1}D_{q^{-2}}^{4}f(q^{-1}x^{2},q^{-1}x^{3})+q^{-1}\lambda
x^{1}D_{q^{-2}}^{2}D_{q^{-2}}^{3}f, \\
\hat{\partial}^{2}\triangleright f &=&D_{q^{-2}}^{3}f(q^{-1}x^{1}), 
\nonumber \\
\hat{\partial}^{3}\triangleright f &=&D_{q^{-2}}^{2}f(q^{-1}x^{1}), 
\nonumber \\
\hat{\partial}^{4}\triangleright f &=&qD_{q^{-2}}^{1}f.  \nonumber
\end{eqnarray}
For the sake of simplicity the representations of the unhated partial
derivatives refer to a different ordering, namely
$X^{4}X^{3}X^{2}X^{1}.$ In this setting they are given by 
\begin{eqnarray}
\partial ^{4}\tilde{\triangleright}f
&=&qD_{q^{2}}^{1}f(qx^{2},qx^{3})-q\lambda x^{4}D_{q^{2}}^{2}D_{q^{2}}^{3}f,
\\
{\partial}^{3}\tilde{\triangleright}f &=&D_{q^{2}}^{2}f(qx^{4}), 
\nonumber \\
{\partial}^{2}\tilde{\triangleright}f &=&D_{q^{2}}^{3}f(qx^{4}), 
\nonumber \\
{\partial}^{1}\tilde{\triangleright}f &=&q^{-1}D_{q^{2}}^{4}f.  \nonumber
\end{eqnarray}
And if we want to have representations belonging to one given ordering
only, we can apply the formulae
\begin{eqnarray}
\hat{\partial}^{i}\triangleright (\hat{U}^{-1}f) &=&\hat{U}^{-1}(\hat{%
\partial}^{i}\tilde{\triangleright}f), \\
{\partial}^{i}\tilde{\triangleright}(\hat{U}f) &=&\hat{U}({\partial}%
^{i}\triangleright f)  \nonumber
\end{eqnarray}
with
\begin{eqnarray}
\hat{U}^{-1}f   
&=&\sum_{i=0}^{\infty }\lambda ^{i}\frac{(x^{2}x^{3})^{i}}{[[i]]_{q^{-2}}!}%
q^{-(\hat{n}_{2}+\hat{n}_{3})\left( \hat{n}_{1}+\hat{n}_{4}+i\right) }\left(
D_{q^{-2}}^{4}D_{q^{-2}}^{1}\right) ^{i}f,
\label{UmordOp4dim} \\%
[0.16in]
\hat{U}f  
&=&\sum_{i=0}^{\infty }\left( -\lambda \right) ^{i}\frac{(x^{2}x^{3})^{i}}{%
[[i]]_{q^{2}}!}q^{(\hat{n}_{2}+\hat{n}_{3})\left( \hat{n}_{1}+\hat{n}%
_{4}+i\right) }\left( D_{q^{2}}^{1}D_{q^{2}}^{4}\right) ^{i}f.
 \label{UmordOp4dimInv} 
\end{eqnarray}
Using the conjugation properties \cite{Oca96} 
\begin{equation}
\overline{X^{1}}=q^{-1}X^{4},\quad \overline{X^{2}}=X^{3},\quad \overline{%
X^{3}}=X^{2},\quad \overline{X^{4}}=qX^{1},
\end{equation}
\[
\overline{\partial ^{1}}=-q^{-5}\hat{\partial}^{4},\quad \overline{\partial
^{2}}=-q^{-4}\hat{\partial}^{3},\quad \overline{\partial ^{3}}=-q^{-4}\hat{%
\partial}^{2},\quad \overline{\partial ^{4}}=-q^{-3}\hat{\partial}^{1},
\]
\[
\overline{L_{i}^{+}}=q^{-2}L_{i}^{-},\quad \overline{L_{i}^{-}}%
=q^{2}L_{i}^{+},\quad i=1,2
\]
and taking the considerations mentioned in the last section we again find
the translation rules 
\begin{eqnarray}
f\triangleleft \partial ^{i}&\stackrel{j\leftrightarrow j^{\prime }}{%
\longleftrightarrow }&-q^{-4}\hat{\partial}^{i^{\prime }}\triangleright f, \\
f\triangleleft \hat{\partial}^{i}&\stackrel{j\leftrightarrow j^{\prime }}{%
\longleftrightarrow }&-q^{4}\partial ^{i^{\prime }}\triangleright f,\quad
i=1,\ldots ,4,\quad i^{\prime }=5-i,  \nonumber \\[0.16in]
f\triangleleft L_{i}^{+}&\stackrel{j\leftrightarrow j^{\prime }}{%
\longleftrightarrow }&q^{-3}L_{i}^{-}\triangleright f,  \nonumber \\
f\triangleleft L_{i}^{-}&\stackrel{j\leftrightarrow j^{\prime }}{%
\longleftrightarrow }&q^{3}L_{i}^{+}\triangleright f,\quad i=1,2.  \nonumber
\end{eqnarray}
The symbol $\stackrel{j\leftrightarrow j^{\prime }}{\longleftrightarrow }$
now indicates that one can make a transition between the two expressions by
the substitution 
\begin{equation}
x^{j}\longleftrightarrow x^{j^{\prime }},\quad
D_{q^{a}}^{j}\longleftrightarrow D_{q^{a}}^{j^{\prime }},\quad
\hat{n}^j \longleftrightarrow \hat{n}^{j'}
\end{equation}
where $j=1,\ldots ,4,$ $j^{\prime }=5-i.$ An example shall illustrate this: 
\begin{equation}
D_{q^{2}}^{1}D_{q^{2}}^{2}f(qx^{1},q^{2}x^{3})\stackrel{j\leftrightarrow
j^{\prime }}{\longleftrightarrow }%
D_{q^{2}}^{4}D_{q^{2}}^{3}f(qx^{4},q^{2}x^{2}).
\end{equation}
Last but not least we have to treat the representations of the diagonal
generators $K_{1},$ $K_{2},$ $\Lambda $, which can be derived from the
identity (\ref{linksdar}) quite easily. Thus we have 
\begin{eqnarray}
f\triangleleft K_{1} &=&(K_{1})^{-1}\triangleright f, \\
f\triangleleft K_{2} &=&(K_{2})^{-1}\triangleright f, \\
f\triangleleft \Lambda ^{\pm \frac{1}{2}} &=&\Lambda ^{\mp \frac{1}{2}%
}\triangleright f.  \nonumber
\end{eqnarray}

\section{q-Deformed Minkowski space\label{chap4}}

From a physical point of view q-deformed Minkowski space \cite{CSSW90}, \cite
{SWZ91}, \cite{Maj91}\footnote{%
For a different version of q-deformed Minkowski space see also \cite{Dob94}.}
is the most interesting one of all considered cases. In addition a treatment
is desirable which pays certain attention to the central time element $X^{0}$
\cite{LSW94}. The general form of the commutation relations between partial
derivatives and space-time coordinates now reads \cite{LWW97} 
\begin{equation}
\partial ^{A}X^{B}=\eta ^{AB}+q^{-2}(\hat{R}_{II}^{-1})_{CD}^{AB}X^{C}%
\partial ^{D},\quad A,B,C,D=0,3,\pm 
\end{equation}
where $\eta ^{AB}$ denotes the metric and $\ \hat{R}_{II}$ one of the two $R$-%
matrices of q-deformed Minkowski space \cite{LSW94}. For the sake of simplicity we
introduce the light cone coordinate $\tilde{X}^{3}=X^{3}-X^{0}$ and
the corresponding
partial derivative $\tilde{\partial}^{3}=\partial ^{3}-\partial ^{0}.$ In
terms of these quantities the above relations become 
\begin{eqnarray}
\tilde{\partial}^{3}\tilde{X}^{3} &=&\tilde{X}^{3}\tilde{\partial}^{3},
\label{AblKoordMinAnf} \\
\tilde{\partial}^{3}X^{+} &=&X^{+}\tilde{\partial}^{3}+q^{-1}\lambda \tilde{X%
}^{3}\partial ^{+},  \nonumber \\
\tilde{\partial}^{3}X^{3} &=&1+q^{-2}X^{3}\tilde{\partial}^{3}+q^{-2}\lambda
\lambda _{+}^{-1}\tilde{X}^{3}\tilde{\partial}^{3}+q^{-2}\lambda
X^{-}\partial ^{+},  \nonumber \\
\tilde{\partial}^{3}X^{-} &=&q^{-2}X^{-}\tilde{\partial}^{3},  \nonumber \\%
[0.16in]
\partial ^{+}\tilde{X}^{3} &=&q^{-2}\tilde{X}^{3}\partial ^{+}, \\
\partial ^{+}X^{+} &=&X^{+}\partial ^{+},  \nonumber \\
\partial ^{+}X^{3} &=&X^{3}\partial ^{+}-\lambda \lambda _{+}^{-1}\tilde{X}%
^{3}\partial ^{+}-\lambda \lambda _{+}^{-1}X^{+}\tilde{\partial}^{3}, 
\nonumber \\
\partial ^{+}X^{-} &=&-q+q^{-2}X^{-}\partial ^{+}-q^{-1}\lambda \lambda
_{+}^{-1}\tilde{X}^{3}\partial ^{3},  \nonumber \\[0.16in]
\partial ^{-}\tilde{X}^{3} &=&\tilde{X}^{3}\partial ^{-}+q^{-1}\lambda X^{-}%
\tilde{\partial}^{3}, \\
\partial ^{-}X^{+} &=&-q^{-1}+q^{-2}X^{+}\partial ^{-}+q^{-2}\lambda \tilde{X%
}^{3}\partial ^{0}+q^{-2}\lambda ^{2}X^{-}\partial ^{+}  \nonumber \\
&&+\,q^{-2}\lambda X^{3}\tilde{\partial}^{3}+q^{-1}\lambda \lambda _{+}^{-1}%
\tilde{X}^{3}\tilde{\partial}^{3},  \nonumber \\
\partial ^{-}X^{3} &=&q^{-2}X^{3}\partial ^{-}+q^{-2}\lambda \lambda
_{+}^{-1}\tilde{X}^{3}\partial ^{-}+q^{-1}\lambda X^{-}\partial ^{0} 
\nonumber \\
&&+\,\lambda \lambda _{+}^{-1}(1+2q^{-2})X^{-}\tilde{\partial}^{3}, 
\nonumber \\
\partial ^{-}X^{-} &=&X^{-}\partial ^{-},  \nonumber \\[0.16in]
\partial ^{0}\tilde{X}^{3} &=&1+q^{-2}\tilde{X}^{3}\partial
^{0}+q^{-2}\lambda X^{-}\partial ^{+}-\lambda \lambda _{+}^{-1}\tilde{X}^{3}%
\tilde{\partial}^{3},  \label{AblKoordMin End} \\
\partial ^{0}X^{+} &=&q^{-2}X^{+}\partial ^{0}+q^{-1}\lambda X^{3}\partial
^{+}-\lambda \lambda _{+}^{-1}\tilde{X}^{3}\partial ^{+}-\lambda \lambda
_{+}^{-1}X^{+}\tilde{\partial}^{3},  \nonumber \\
\partial ^{0}X^{3} &=&X^{3}\partial ^{0}-\lambda \lambda _{+}^{-1}\tilde{X}%
^{3}\partial ^{0}-q^{-1}\lambda \lambda _{+}^{-1}X^{+}\partial ^{-} 
\nonumber \\
&&+\,q^{-1}\lambda \lambda _{+}^{-1}X^{-}\partial ^{+}+q^{-2}\lambda \lambda
_{+}^{-1}X^{3}\tilde{\partial}^{3}-\lambda \lambda _{+}^{-1}\tilde{X}^{3}%
\tilde{\partial}^{3},  \nonumber \\
\partial ^{0}X^{-} &=&X^{-}\partial ^{0}+q^{-2}\lambda \lambda _{+}^{-1}X^{-}%
\tilde{\partial}^{3}-\lambda \lambda _{+}^{-1}\tilde{X}^{3}\partial ^{-}. 
\nonumber
\end{eqnarray}
For the second differential calculus we have the relation 
\begin{equation}
\hat{\partial}^{A}X^{B}=\eta ^{AB}+q^{2}(\hat{R}_{II})_{CD}^{AB}X^{C}\hat{%
\partial}^{D},
\end{equation}
which gives in a more explicit form 
\begin{eqnarray}
\hat{\tilde{\partial}^{3}}\tilde{X}^{3} &=&\tilde{X}^{3}\hat{\tilde{\partial}%
^{3}},  \label{D3Min} \\
\hat{\tilde{\partial}^{3}}X^{-} &=&X^{-}\hat{\tilde{\partial}^{3}}-q\lambda 
\tilde{X}^{3}\hat{\partial}^{-},  \nonumber \\
\hat{\tilde{\partial}^{3}}X^{3} &=&1+q^{2}X^{3}\hat{\tilde{\partial}^{3}}%
-q^{2}\lambda \lambda _{+}^{-1}\tilde{X}^{3}\hat{\tilde{\partial}^{3}}%
-q^{2}\lambda X^{+}\hat{\partial}^{-},  \nonumber \\
\hat{\tilde{\partial}^{3}}X^{+} &=&q^{2}X^{+}\hat{\tilde{\partial}^{3}}, 
\nonumber \\[0.16in]
\hat{\partial}^{-}\tilde{X}^{3} &=&q^{2}\tilde{X}^{3}\hat{\partial}^{-},
\label{D+Min} \\
\hat{\partial}^{-}X^{-} &=&X^{-}\hat{\partial}^{-},  \nonumber \\
\hat{\partial}^{-}X^{3} &=&X^{3}\hat{\partial}^{-}+\lambda \lambda _{+}^{-1}%
\tilde{X}^{3}\hat{\partial}^{-}+\lambda \lambda _{+}^{-1}X^{-}\hat{\tilde{%
\partial}^{3}},  \nonumber \\
\hat{\partial}^{-}X^{+} &=&-q^{-1}+q^{2}X^{+}\hat{\partial}^{-}+q\lambda
\lambda _{+}^{-1}\tilde{X}^{3}\hat{\tilde{\partial}^{3}},  \nonumber \\%
[0.16in]
\hat{\partial}^{+}\tilde{X}^{3} &=&\tilde{X}^{3}\hat{\partial}^{+}-q\lambda
X^{+}\hat{\tilde{\partial}^{3}}, \\
\hat{\partial}^{+}X^{-} &=&-q+q^{2}X^{-}\hat{\partial}^{+}-q^{2}\lambda 
\tilde{X}^{3}\hat{\partial}^{0}+q^{2}\lambda ^{2}X^{+}\hat{\partial}^{-} 
\nonumber \\
&&-\,q^{2}\lambda X^{3}\hat{\tilde{\partial}^{3}}-q\lambda \lambda _{+}^{-1}%
\tilde{X}^{3}\hat{\tilde{\partial}^{3}},  \nonumber \\
\hat{\partial}^{+}X^{3} &=&q^{2}X^{3}\hat{\partial}^{+}-q^{2}\lambda \lambda
_{+}^{-1}\tilde{X}^{3}\hat{\partial}^{+}-q\lambda X^{+}\hat{\partial}^{0} 
\nonumber \\
&&-\,\lambda \lambda _{+}^{-1}(1+2q^{2})X^{+}\hat{\tilde{\partial}^{3}}, 
\nonumber \\
\hat{\partial}^{+}X^{+} &=&X^{+}\hat{\partial}^{+},  \nonumber \\[0.16in]
\hat{\partial}^{0}\tilde{X}^{3} &=&1+q^{2}\tilde{X}^{3}\hat{\partial}%
^{0}-q^{2}\lambda X^{+}\hat{\partial}^{-}+\lambda \lambda _{+}^{-1}\tilde{X}%
^{3}\hat{\tilde{\partial}^{3}},  \label{DEnde} \\
\hat{\partial}^{0}X^{-} &=&q^{2}X^{-}\hat{\partial}^{0}-q\lambda X^{3}\hat{%
\partial}^{-}+\lambda \lambda _{+}^{-1}\tilde{X}^{3}\hat{\partial}^{-}+\lambda
\lambda _{+}^{-1}X^{-}\hat{\tilde{\partial}^{3}},  \nonumber \\
\hat{\partial}^{0}X^{3} &=&X^{3}\hat{\partial}^{0}+\lambda \lambda _{+}^{-1}%
\tilde{X}^{3}\hat{\partial}^{0}+q\lambda \lambda _{+}^{-1}X^{-}\hat{\partial}%
^{+}-q\lambda \lambda _{+}^{-1}X^{+}\hat{\partial}^{-}  \nonumber \\
&&-\,q^{2}\lambda \lambda _{+}^{-1}X^{3}\hat{\tilde{\partial}^{3}}+\lambda
\lambda _{+}^{-1}\tilde{X}^{3}\hat{\tilde{\partial}^{3}},  \nonumber \\
\hat{\partial}^{0}X^{+} &=&X^{+}\hat{\partial}^{0}-q^{2}\lambda \lambda
_{+}^{-1}X^{+}\hat{\tilde{\partial}^{3}}+\lambda \lambda _{+}^{-1}\tilde{X}%
^{3}\hat{\partial}^{+}.  \nonumber
\end{eqnarray}
From these relations we can again deduce a Hopf structure \cite{OSWZ92}. In
terms of Lorentz generators $T^{+},$ $T^{-},$ $\tau ^{3},$ $T^{2},$ $S^{1},$
\ $\tau ^{1},$ $\sigma ^{2}$ and the scaling operator $\Lambda $ the Hopf
structure of the derivatives $\partial ^{0},$ $\partial ^{+},$ $\partial
^{-},$ $\tilde{\partial}^{3}$ becomes 
\begin{eqnarray}
\Delta (\tilde{\partial}^{3}) &=&\tilde{\partial}^{3}\otimes 1+\Lambda ^{%
\frac{1}{2}}\tau ^{1}\otimes \tilde{\partial}^{3}-q^{\frac{1}{2}}\lambda
_{+}^{\frac{1}{2}}\lambda \Lambda ^{\frac{1}{2}}(\tau ^{3})^{-\frac{1}{2}%
}S^{1}\otimes \partial ^{+}, \\
\Delta (\partial ^{+}) &=&\partial ^{+}\otimes 1+\Lambda ^{\frac{1}{2}}(\tau
^{3})^{-\frac{1}{2}}\sigma ^{2}\otimes \partial ^{+}-q^{\frac{3}{2}}\lambda
_{+}^{-\frac{1}{2}}\lambda \Lambda ^{\frac{1}{2}}T^{2}\otimes \tilde{\partial%
}^{3},  \nonumber \\
\Delta (\partial ^{-}) &=&\partial ^{-}\otimes 1+\Lambda ^{\frac{1}{2}}(\tau
^{3})^{\frac{1}{2}}\tau ^{1}\otimes \partial ^{-}-q^{-\frac{1}{2}}\lambda
_{+}^{\frac{1}{2}}\lambda \Lambda ^{\frac{1}{2}}S^{1}\otimes \partial ^{0} 
\nonumber \\
&&-\,\lambda ^{2}\Lambda ^{\frac{1}{2}}(\tau ^{3})^{-\frac{1}{2}%
}T^{-}S^{1}\otimes \partial ^{+}  \nonumber \\
&&+\,q^{-\frac{1}{2}}\lambda _{+}^{-\frac{1}{2}}\lambda \Lambda ^{\frac{1}{2}%
}(\tau ^{1}T^{-}-q^{-1}S^{1})\otimes \tilde{\partial}^{3},  \nonumber \\
\Delta (\partial ^{0}) &=&\partial ^{0}\otimes 1+\Lambda ^{\frac{1}{2}%
}\sigma ^{2}\otimes \partial ^{0}-q^{\frac{1}{2}}\lambda _{+}^{-\frac{1}{2}%
}\lambda \Lambda ^{\frac{1}{2}}T^{2}(\tau ^{3})^{\frac{1}{2}}\otimes
\partial ^{-}  \nonumber \\
&&+\,q^{\frac{1}{2}}\lambda _{+}^{-\frac{1}{2}}\lambda \Lambda ^{\frac{1}{2}%
}\left( \tau ^{3}\right) ^{-\frac{1}{2}}(T^{-}\sigma ^{2}+qS^{1})\otimes
\partial ^{+}  \nonumber \\
&&-\,\lambda _{+}^{-1}\Lambda ^{\frac{1}{2}}(\lambda ^{2}T^{-}T^{2}+q(\tau
^{1}-\sigma ^{2}))\otimes \tilde{\partial}^{3},  \nonumber \\[0.16in]
S(\tilde{\partial}^{3}) &=&-\Lambda ^{-\frac{1}{2}}\sigma ^{2}\tilde{\partial%
}^{3}-q^{-\frac{3}{2}}\lambda _{+}^{\frac{1}{2}}\lambda \Lambda ^{-\frac{1}{2%
}}S^{1}\partial ^{+}, \\
S(\partial ^{+}) &=&-\Lambda ^{-\frac{1}{2}}\tau ^{1}(\tau ^{3})^{\frac{1}{2}%
}\partial ^{+}-q^{\frac{3}{2}}\lambda _{+}^{-\frac{1}{2}}\lambda \Lambda ^{-%
\frac{1}{2}}T^{2}(\tau ^{3})^{\frac{1}{2}}\tilde{\partial}^{3},  \nonumber \\
S(\partial ^{-}) &=&-\Lambda ^{-\frac{1}{2}}\sigma ^{2}(\tau ^{3})^{-\frac{1%
}{2}}\partial ^{-}-q^{-\frac{1}{2}}\lambda _{+}^{\frac{1}{2}}\lambda \Lambda
^{-\frac{1}{2}}(\tau ^{3})^{-\frac{1}{2}}S^{1}\partial ^{0}  \nonumber \\
&&+\,q^{-2}\lambda ^{2}\Lambda ^{-\frac{1}{2}}(\tau ^{3})^{-\frac{1}{2}%
}S^{1}T^{-}\partial ^{+}  \nonumber \\
&&+\,q^{-\frac{5}{2}}\lambda _{+}^{-\frac{1}{2}}\lambda \Lambda ^{-\frac{1}{2%
}}(\tau ^{3})^{-\frac{1}{2}}(\sigma ^{2}T^{-}-q^{3}S)\tilde{\partial}^{3}, 
\nonumber \\
S(\partial ^{0}) &=&-\Lambda ^{-\frac{1}{2}}\tau ^{1}\partial ^{0}-q^{\frac{5%
}{2}}\lambda _{+}^{-\frac{1}{2}}\lambda \Lambda ^{-\frac{1}{2}}T^{2}\partial
^{-}  \nonumber \\
&&+\,q^{-\frac{3}{2}}\lambda _{+}^{-\frac{1}{2}}\lambda \Lambda ^{-\frac{1}{2%
}}(\tau ^{1}T^{-}+qS^{1})\partial ^{+}  \nonumber \\
&&+\,\lambda _{+}^{-1}\Lambda ^{-\frac{1}{2}}(q(\sigma ^{2}-\tau
^{1})+\lambda ^{2}T^{2}T^{-})\tilde{\partial}^{3},  \nonumber \\[0.16in]
\varepsilon (\tilde{\partial}^{3}) &=&\varepsilon (\partial
^{+})=\varepsilon (\partial ^{-})=\varepsilon (\partial ^{0})=0.
\end{eqnarray}
Similar expressions can be found for the second set of derivatives: 
\begin{eqnarray}
\Delta \hat{(\tilde{\partial}^{3})} &=&\hat{\tilde{\partial}^{3}}\otimes
1+\Lambda ^{-\frac{1}{2}}\left( \tau ^{3}\right) ^{-\frac{1}{2}}\sigma
^{2}\otimes \hat{\tilde{\partial}^{3}}-q^{\frac{3}{2}}\lambda _{+}^{\frac{1}{%
2}}\lambda \Lambda ^{-\frac{1}{2}}T^{2}\otimes \hat{\partial}^{-},  \nonumber
\\
\Delta (\hat{\partial}^{-}) &=&\hat{\partial}^{-}\otimes 1+\Lambda ^{-\frac{1%
}{2}}\tau ^{1}\otimes \hat{\partial}^{-}-q^{\frac{1}{2}}\lambda _{+}^{-\frac{%
1}{2}}\lambda \Lambda ^{-\frac{1}{2}}(\tau ^{3})^{-\frac{1}{2}}S^{1}\otimes 
\hat{\tilde{\partial}^{3}},  \nonumber \\
\Delta (\hat{\partial}^{+}) &=&\hat{\partial}^{+}\otimes 1+\Lambda ^{-\frac{1%
}{2}}\sigma ^{2}\otimes \hat{\partial}^{+}-q^{\frac{1}{2}}\lambda _{+}^{%
\frac{1}{2}}\lambda \Lambda ^{-\frac{1}{2}}T^{2}(\tau ^{3})^{\frac{1}{2}%
}\otimes \hat{\partial}^{0}  \nonumber \\
&&-\,q^{\frac{1}{2}}\lambda _{+}^{-\frac{1}{2}}\lambda \Lambda ^{-\frac{1}{2}%
}(\tau ^{3})^{-\frac{1}{2}}(T^{+}\sigma ^{2}+q\tau ^{3}T^{2})\otimes \hat{%
\tilde{\partial}^{3}}  \nonumber \\
&&+\,q^{2}\lambda ^{2}\Lambda ^{-\frac{1}{2}}T^{2}T^{+}\otimes \hat{\partial}%
^{-},  \nonumber \\
\Delta (\hat{\partial}^{0}) &=&\hat{\partial}^{0}\otimes 1+\Lambda ^{-\frac{1%
}{2}}(\tau ^{3})^{\frac{1}{2}}\tau ^{1}\otimes \hat{\partial}^{0}-q^{-\frac{%
1}{2}}\lambda _{+}^{-\frac{1}{2}}\lambda \Lambda ^{-\frac{1}{2}}S^{1}\otimes 
\hat{\partial}^{+}  \nonumber \\
&&-\,q^{\frac{1}{2}}\lambda _{+}^{-\frac{1}{2}}\lambda \Lambda ^{-\frac{1}{2}%
}(qT^{+}\tau ^{1}-T^{2})\otimes \hat{\partial}^{-}  \nonumber \\
&&+\,\lambda _{+}^{-1}\Lambda ^{-\frac{1}{2}}(\tau ^{3})^{-\frac{1}{2}%
}(\lambda ^{2}T^{+}S^{1}+q^{-1}(\tau ^{3}\tau ^{1}-\sigma ^{2}))\otimes \hat{%
\tilde{\partial}^{3}},  \nonumber \\[0.16in]
S(\hat{\tilde{\partial}^{3})} &=&-\Lambda ^{\frac{1}{2}}\tau ^{1}(\tau
^{3})^{\frac{1}{2}}\hat{\tilde{\partial}^{3}}-q^{\frac{3}{2}}\lambda _{+}^{%
\frac{1}{2}}\lambda \Lambda ^{\frac{1}{2}}T^{2}(\tau ^{3})^{\frac{1}{2}}\hat{%
\partial}^{-}, \\
S(\hat{\partial}^{-}) &=&-\Lambda ^{\frac{1}{2}}\sigma ^{2}\hat{\partial}%
^{-}-q^{-\frac{3}{2}}\lambda _{+}^{-\frac{1}{2}}\lambda \Lambda ^{\frac{1}{2}%
}S^{1}\hat{\tilde{\partial}^{3}},  \nonumber \\
S(\hat{\partial}^{+}) &=&-\Lambda ^{\frac{1}{2}}\tau ^{1}\hat{\partial}%
^{+}-q^{\frac{5}{2}}\lambda _{+}^{\frac{1}{2}}\lambda \Lambda ^{\frac{1}{2}%
}T^{2}\hat{\partial}^{0}  \nonumber \\
&&-\,q^{\frac{3}{2}}\lambda _{+}^{-\frac{1}{2}}\lambda \Lambda ^{\frac{1}{2}%
}(q\tau ^{1}T^{+}+T^{2})\hat{\tilde{\partial}^{3}}  \nonumber \\
&&-\,q^{4}\lambda ^{2}\Lambda ^{\frac{1}{2}}T^{2}T^{+}\hat{\partial}^{-}, 
\nonumber \\
S(\hat{\partial}^{0}) &=&-\Lambda ^{\frac{1}{2}}(\tau ^{3})^{-\frac{1}{2}%
}\sigma ^{2}\hat{\partial}^{0}-q^{-\frac{1}{2}}\lambda _{+}^{-\frac{1}{2}%
}\lambda \Lambda ^{\frac{1}{2}}(\tau ^{3})^{-\frac{1}{2}}S^{1}\hat{\partial}%
^{+}  \nonumber \\
&&-\,q^{\frac{3}{2}}\lambda _{+}^{-\frac{1}{2}}\lambda \Lambda ^{\frac{1}{2}%
}(\tau ^{3})^{-\frac{1}{2}}(\sigma ^{2}T^{+}-q\tau ^{3}T^{2})\hat{\partial}%
^{-}  \nonumber \\
&&-\,\lambda _{+}^{-1}\Lambda ^{\frac{1}{2}}(\tau ^{3})^{-\frac{1}{2}%
}(\lambda ^{2}T^{+}S^{1}+q(\sigma ^{2}-\tau ^{3}\tau ^{1}))\hat{\tilde{%
\partial}^{3}},  \nonumber \\[0.16in]
\varepsilon \hat{(\tilde{\partial}^{3})} &=&\varepsilon (\hat{\partial}%
^{+})=\varepsilon (\hat{\partial}^{-})=\varepsilon (\hat{\partial}^{0})=0. 
\nonumber
\end{eqnarray}
The Leibniz rules can be read off from the formulae for the co-product as
usual, if one knows the representations \ of the given Lorentz generators
and the scaling operator $\Lambda $. These representations, however, can be
obtained from the commutation relations \cite{OSWZ92} 
\begin{eqnarray}
T^{+}X^{0} &=&X^{0}T^{+}, \\
T^{+}\tilde{X}^{3} &=&\tilde{X}^{3}T^{+}+q^{-\frac{3}{2}}\lambda _{+}^{\frac{%
1}{2}}X^{+},  \nonumber \\
T^{+}X^{+} &=&q^{-2}X^{+}T^{+},  \nonumber \\
T^{+}X^{-} &=&q^{2}X^{-}T^{+}+q^{-\frac{1}{2}}\lambda _{+}^{\frac{1}{2}%
}X^{3},  \nonumber \\[0.16in]
T^{-}X^{0} &=&X^{0}T^{-}, \\
T^{-}\tilde{X}^{3} &=&\tilde{X}^{3}T^{-}+q^{\frac{3}{2}}\lambda _{+}^{\frac{1%
}{2}}X^{-},  \nonumber \\
T^{-}X^{-} &=&q^{2}X^{-}T^{-},  \nonumber \\
T^{-}X^{+} &=&q^{-2}X^{+}T^{-}+q^{\frac{1}{2}}\lambda _{+}^{\frac{1}{2}%
}X^{3},  \nonumber \\[0.16in]
\tau ^{3}X^{0} &=&X^{0}\tau ^{3}, \\
\tau ^{3}\tilde{X}^{3} &=&\tilde{X}^{3}\tau ^{3},  \nonumber \\
\tau ^{3}X^{+} &=&q^{-4}X^{+}\tau ^{3},  \nonumber \\
\tau ^{3}X^{-} &=&q^{4}X^{-}\tau ^{3},  \nonumber \\[0.16in]
T^{2}\tilde{X}^{3} &=&q^{-1}\tilde{X}^{3}T^{2}, \\
T^{2}X^{+} &=&qX^{+}T^{2},  \nonumber \\
T^{2}X^{-} &=&q^{-1}X^{-}T^{2}+q^{-\frac{3}{2}}\lambda _{+}^{-\frac{1}{2}}%
\tilde{X}^{3}\tau ^{1},  \nonumber \\
T^{2}X^{3} &=&qX^{3}T^{2}-q\lambda _{+}^{-1}\lambda \tilde{X}^{3}T^{2}+q^{-%
\frac{1}{2}}\lambda _{+}^{\frac{1}{2}}X^{+}\tau ^{1},  \nonumber \\[0.16in]
S^{1}\tilde{X}^{3} &=&q\tilde{X}^{3}S^{1}, \\
S^{1}X^{-} &=&qX^{-}S^{1},  \nonumber \\
S^{1}X^{+} &=&q^{-1}X^{+}S^{1}-q^{-\frac{1}{2}}\lambda _{+}^{-\frac{1}{2}}%
\tilde{X}^{3}\sigma ^{2},  \nonumber \\
S^{1}X^{3} &=&q^{-1}X^{3}S^{1}+q^{-1}\lambda _{+}^{-1}\lambda \tilde{X}%
^{3}S^{1}-q^{\frac{1}{2}}\lambda _{+}^{-\frac{1}{2}}X^{-}\sigma ^{2}, 
\nonumber \\[0.16in]
\tau ^{1}\tilde{X}^{3} &=&q\tilde{X}^{3}\tau ^{1}, \\
\tau ^{1}X^{-} &=&q^{-1}X^{-}\tau ^{1},  \nonumber \\
\tau ^{1}X^{+} &=&qX^{+}\tau ^{1}-q^{\frac{3}{2}}\lambda _{+}^{-\frac{1}{2}%
}\lambda ^{2}\tilde{X}^{3}T^{2},  \nonumber \\
\tau ^{1}X^{3} &=&q^{-1}X^{3}\tau ^{1}+q^{-1}\lambda _{+}^{-1}\lambda \tilde{%
X}^{3}\tau ^{1}-q^{\frac{1}{2}}\lambda _{+}^{-\frac{1}{2}}\lambda
^{2}X^{-}T^{2},  \nonumber \\[0.16in]
\sigma ^{2}\tilde{X}^{3} &=&q^{-1}\tilde{X}^{3}\sigma ^{2}, \\
\sigma ^{2}X^{+} &=&q^{-1}X^{+}\sigma ^{2},  \nonumber \\
\sigma ^{2}X^{-} &=&qX^{-}\sigma ^{2}+q^{\frac{1}{2}}\lambda _{+}^{-\frac{1}{%
2}}\lambda ^{2}\tilde{X}^{3}S^{1},  \nonumber \\
\sigma ^{2}X^{3} &=&qX^{3}\sigma ^{2}-q\lambda _{+}^{-1}\lambda \tilde{X}%
^{3}\sigma ^{2}+q^{-\frac{1}{2}}\lambda _{+}^{-\frac{1}{2}}\lambda
^{2}X^{+}S^{1},  \nonumber \\[0.16in]
\Lambda X^{A} &=&q^{-2}X^{A}\Lambda ,\qquad A=0,3,\pm .
\end{eqnarray}

To calculate representations for partial derivatives and Lorentz generators
we need to take special considerations into account. We want to demonstrate
this by a short example. By multiple use of the relations (\ref{D3Min}-\ref
{D+Min}), one can show that the following identity holds 
\begin{eqnarray}
&&\tilde{\partial}^{3}(X^{3})^{n}= \label{BeispDreiAblMin} \\
&&\sum_{k=0}^{\left[ \frac{n}{2}\right] }\Big(q^{-2}\frac{\lambda
^{2}}{\lambda
_{+}}\Big)^{k}\sum_{j_{1}=0}^{n-2k}\sum_{j_{2}=0}^{n-2k-j_{1}}\ldots
\sum_{j_{2k}=0}^{n-2k-j_{1}-\ldots -j_{2k-1}}q^{2(j_{2}+j_{4}+\ldots
+j_{2k})}  \nonumber \\
&&\hspace{1in}\cdot\; q^{-2(n-2k)}(X^{3}+\lambda \lambda _{+}^{-1}\tilde{X}%
^{3})^{n-2k}(X^{-}X^{+})^{k}\tilde{\partial}^{3}  \nonumber \\
&&+\;{}q^{-2}\lambda \sum_{k=0}^{\left[ \frac{n-1}{2}\right]
}\Big(q^{-2}\frac{\lambda ^{2}}{\lambda _{+}}\Big)^{k}
\sum_{j_{1}=0}^{n-2k-1}\ldots
\sum_{j_{2k+1}=0}^{n-2k-1-j_{1}-\ldots -j_{2k}}q^{-2(j_{1}+j_{3}+\ldots
+j_{2k+1})}  \nonumber \\
&&\hspace{1in}\cdot (X^{3}+\lambda \lambda _{+}^{-1}\tilde{X}%
^{3})^{n-2k-1}(X^{-}X^{+})^{k}X^{-}\partial ^{+}  \nonumber \\
&&+\sum_{k=0}^{\left[ \frac{n-1}{2}\right] }\Big(q^{-2}\frac{\lambda
^{2}}{\lambda _{+}}\Big)^{k}\sum_{j_{1}=0}^{n-2k-1}\ldots
\sum_{j_{2k+1}=0}^{n-2k-1-j_{1}-\ldots -j_{2k}}q^{-2(j_{1}+j_{3}+\ldots
+j_{2k+1})}  \nonumber \\
&&\hspace{1in}\cdot (X^{3}+\lambda \lambda _{+}^{-1}\tilde{X}%
^{3})^{j_{1}+j_{2}+\ldots j_{2k+1}}  \nonumber \\
&&\hspace{1in}\cdot (X^{-}X^{+})^{k}(X^{3})^{n-2k-1-j_{1}-\ldots -j_{2k+1}} 
\nonumber
\end{eqnarray}
where $[s]$ denotes the biggest integer not being bigger than s. To go
further, one has to overcome two difficulties. The first one is to give
normal ordered expressions for $(X^{+}X^{-})^{k}$ and $(X^{-}X^{+})^{k}.$
Towards this aim we start from the relations 
\begin{eqnarray}
\hat{r}^{2} &=&-a_{q}(X^{0},\tilde{X}^{3})+\lambda _{+}X^{-}X^{+},\\
\hat{r}^{2} &=&-a_{q^{-1}}(X^{0},\tilde{X}^{3})+\lambda_{+}X^{+}X^{-},\nonumber
\end{eqnarray}
where 
\begin{equation}
a_{q}(X^{0},\tilde{X}^{3}) =q^{2}(\tilde{X}^{3})^{2}+q\lambda _{+}X^{0}%
\tilde{X}^{3} 
\end{equation}
and solve for $X^{-}X^{+}$ and $X^{+}X^{-}$. Thus we can write 
\begin{eqnarray}
\hspace{-0.25in}(X^{-}X^{+})^{k} &=&(\lambda _{+})^{-k}(\hat{r}%
^{2}+a_{q}(X^{0},\tilde{X}^{3}))^{k}  \label{X-X+} \\
&=&(\lambda _{+})^{-k}\sum_{i=0}^{k}\binom{k}{i}\hat{r}^{2i}(a_{q}(X^{0},%
\tilde{X}^{3}))^{k-i}  \nonumber \\
&=&(\lambda _{+})^{-k}\sum_{i=0}^{k}\binom{k}{i}\sum_{p=0}^{i}\lambda
_{+}^{p}(X^{+})^{p}(a_{q}(X^{0},q^{2p}\tilde{X}^{3}))^{k-i}  \nonumber \\
&&\hspace{1.4in}\cdot (S_{q})_{i,p}(X^{0},\tilde{X}^{3})(X^{-})^{p}, 
\nonumber \\[0.16in]
\hspace{-0.25in}(X^{+}X^{-})^{k} &=&(\lambda _{+})^{-k}\sum_{i=0}^{k}\binom{k%
}{i}\sum_{p=0}^{i}\lambda _{+}^{p}(X^{+})^{p}(a_{q^{-1}}(X^{0},q^{2p}\tilde{X}%
^{3}))^{k-i} \\
&&\hspace{1.4in}\cdot (S_{q})_{i,p}(X^{0},\tilde{X}^{3})(X^{-})^{p} 
\nonumber
\end{eqnarray}
where 
\begin{eqnarray}
&&(S_{q})_{k,v}(x^{0},\tilde{x}^{3})=  \label{Flend} \\
&&\left\{ 
\begin{array}{c@{\quad,\quad}l}
1&\text{ if }v=k, \\ 
\sum_{j_{1=0}}^{v}\sum_{j_{2=0}}^{j_{1}}\ldots
\sum_{j_{k-v=0}}^{j_{k-v-1}}\prod_{l=1}^{k-v}a_{q}(q^{2j_{l}}\tilde{x}^{3})&%
\text{ if 0}\leq v<k.
\end{array}
\right.   \nonumber
\end{eqnarray}
For the second equality in (\ref{X-X+}) we have used that $\hat{r}^{2}$ and $%
a_{\pm }$ commute. And for the third equality in (\ref{X-X+}) we have
inserted the normal ordered expression for powers of $\hat{r}^{2}$ which has
been taken from \cite{WW01}.

The second problem we have to adress has to do with the question how can we
generalize our representations to arbitrary functions. As opposed to the
Euclidean cases we cannot rewrite the recursive sums in formula (\ref
{BeispDreiAblMin}) in terms of q-numbers only. However, it should be rather
obvious that these recursive sums can be identified with the following
quantities 
\begin{equation}
(K_{n})_{a_{1},\ldots ,a_{l}}^{(k_{1},\ldots ,k_{l})}\equiv
(K_{n})_{a_{1}}^{(k_{1})}\circ (K_{n-k_{1}})_{a_{2}}^{(k_{2})}\circ \ldots
\circ (K_{n-k_{1}-\ldots -k_{l}})_{a_{l}}^{(k_{l})}
\end{equation}
where
\begin{equation}
(K_{n})_{a}^{(k)}\equiv \sum_{j_{1}=0}^{n-k}\sum_{j_{2}=0}^{n-k-j_{1}}\ldots
\sum_{j_{k}=0}^{n-k-j_{1}-\ldots -j_{k-1}}a^{j_{1}+j_{2}+\ldots
+j_{k}},\quad n\geq k\geq 1,
\end{equation}
and
\begin{equation}
(K_{n})_{a}^{(k)}\circ (K_{n-k})_{b}^{(l)}\equiv
\sum_{j_{1}=0}^{n-k-l}\ldots \sum_{j_{k+l}=0}^{n-k-l-j_{1}-\ldots
-j_{k+l-1}}a^{j_{1}+\ldots +j_{k}}\;b^{j_{k+1}+\ldots +j_{k+l}}.
\end{equation}
Now, the point is that these quantities can be used for defining new linear
operators, if we require for powers of $x$ to hold 
\[
D_{a_{1},\ldots ,a_{l}}^{(k_{1},\ldots ,k_{l})}x^{n}=
(K_{n})_{a_{1},\ldots ,a_{l}}^{(k_{1},\ldots ,k_{l})}x^{n-k_{1}-\ldots
-k_{l}}.
\]
Thus, it remains to derive general formulae for the action of these
operators, a problem which is covered in appendix \ref{AppB}.

In principle, we have everything together for writing down
representations of partial derivatives and Lorentz generators. Before
doing this let us collect some notation that will be used in the
following. First of all we abbreviate 
\begin{eqnarray}
(D_{1,q}^{3})^{k,l} &\equiv &D_{1,q^{2}}^{(k,l)}, \\
(D_{2,q}^{3})^{k,l} &\equiv &D_{y_{-}/x^{3},q^{2}y_{-}/x^{3}}^{(k,l)}, 
\nonumber \\
(D_{3,q}^{3})_{i,j}^{k,l} &\equiv
&D_{y_{+}/x^{3},q^{2}y_{+}/x^{3},y_{-}/x^{3},q^{2}y_{-}/x^{3}}^{(k,l,i,j)} 
\nonumber
\end{eqnarray}
where 
\begin{equation}
y_{\pm }=y_{\pm }(x^{0},\tilde{x}^{3})=x^{0}+\frac{2q^{\pm 1}}{\lambda _{+}}%
\tilde{x}^{3}.
\end{equation}
It should be clear that these operators have to act on the coordinate $x^{3}$
only. Additionally, we make the following definitions 
\begin{eqnarray}
( M^{\pm }) _{i,j}^{k}(\underline{x}) &\equiv &( M^{\pm
}) _{i,j}^{k}(x^{0},x^{+},\tilde{x}^{3},x^{-}) \\
&=&\binom{k}{i}\lambda _{+}^{j}\left( a_{{q}^{\pm 1} }(q^{2j}\tilde{x}^{3})\right)
^{i}\left( x^{+}x^{-}\right) ^{j}(S_q)_{k-j,j}(x^{0},\tilde{x}^{3}),  \nonumber
\\
(M^{+-})_{i,j,u}^{k,l}(\underline{x}) &\equiv
&(M^{+-})_{i,j,u}^{k,l}(x^{0},x^{+},\tilde{x}^{3},x^{-}) \\
&=&\binom{k}{i}\binom{l}{j}\lambda _{+}^{u}\left( a_{^{q^{-1}}}(q^{2u}\tilde{x}%
^{3})\right) ^{l-j}\left( a_{q}(q^{2u}\tilde{x}^{3})\right) ^{k-i}  \nonumber
\\
&&\cdot \left( x^{+}x^{-}\right) ^{u}(S_q)_{i+j,u}(x^{0},\tilde{x}^{3}). 
\nonumber
\end{eqnarray}

Notice that in what follows the normal ordering for which our formulae shall
work is indicated by the sequence of coordinates the given functions depend
on. In this way the representations of the conjugated partial derivatives
become 
\begin{eqnarray}
&&\hat{\tilde{\partial}^{3}}\triangleright f(x^{+},\tilde{x}^{3},x^{3},x^{-})=
\\
&&\sum_{k=0}^{\infty }\alpha _{+}^{k}\sum_{0\leq i+j\leq
k}(M^{-})_{i,j}^{k}(\underline{x})(\tilde{T}^{3})_{j}^{i}f,  \nonumber \\%
[0.16in]
&&\hat{\partial}^{-}\triangleright f(x^{+},\tilde{x}^{3},x^{3},x^{-})= 
-q^{-1}D_{q^{2}}^{+}f  \nonumber \\
&&+{}\frac{\lambda }{\lambda _{+}}\sum_{k=0}^{\infty }\alpha
_{+}^{k}\sum_{0\leq i+j\leq k}\Big\{ (M^{+})_{i,j}^{k}(\underline{x}%
)(T_{1}^{-})_{j}^{i}f+q^{-1}(M^{-})_{i,j}^{k}(\underline{x}%
)(T_{2}^{-})_{j}^{i}f\Big\} ,  \nonumber \\[0.16in]
&&\hat{\partial}^{+}\triangleright f(x^{+},\tilde{x}^{3},x^{3},x^{-})= \\
&&-q\sum_{k=0}^{\infty }\alpha _{+}^{k}\sum_{0\leq i+j\leq k}\Big\{
(M^{-})_{i,j}^{k}(\underline{x})(T_{1}^{+})_{j}^{i}f+\lambda
(M^{+})_{i,j}^{k}(\underline{x})(T_{2}^{+})_{j}^{i}f\Big\}   \nonumber \\
&&-{}q\frac{\lambda }{\lambda _{+}}\sum_{0\leq k+l<\infty }\alpha
_{+}^{k+l}\sum_{i=0}^{k}\sum_{j=0}^{l}\sum_{0\leq u\leq
i+j}(M^{+-})_{i,j,u}^{k,l}(\underline{x})(T_{3}^{+})_{u}^{k,l}f  \nonumber \\
&&-{}\frac{\lambda }{\lambda _{+}}\sum_{0\leq k+l<\infty }\alpha
_{+}^{k+l+1}\sum_{i=0}^{k}\sum_{j=0}^{l+1}\sum_{0\leq u\leq
i+j}(M^{+-})_{i,j,u}^{k,l+1}(\underline{x})(T_{4}^{+})_{u}^{k,l}f,  \nonumber
\\[0.16in]
&&\hat{\partial}^{0}\triangleright f(x^{+},\tilde{x}^{3},x^{3},x^{-})= \\
&&\sum_{k=0}^{\infty }\alpha _{+}^{k}\sum_{0\leq i+j\leq k}\Big\{
(M^{+})_{i,j}^{k}(\underline{x})(T_{1}^{0})_{j}^{i}f-q\frac{\lambda }{%
\lambda _{+}}(M^{-})_{i,j}^{k}(\underline{x})(T_{2}^{0})_{j}^{i}f\Big\}  
\nonumber \\
&&-{}q^{2}\frac{\lambda }{\lambda _{+}}\sum_{0\leq k+l<\infty }\alpha
_{+}^{k+l}\sum_{i=0}^{k}\sum_{j=0}^{l}\sum_{0\leq u\leq
i+j}(M^{+-})_{i,j,u}^{k,l}(\underline{x})(T_{3}^{0})_{u}^{k,l}f
\nonumber \\
&&+{}\frac{\beta }{\lambda _{+}}\sum_{0\leq k+l<\infty }\alpha
_{+}^{k+l+1}\sum_{i=0}^{k+1}\sum_{j=0}^{l}\sum_{0\leq u\leq
i+j}(M^{+-})_{i,j,u}^{k+1,l}(\underline{x})(T_{4}^{0})_{u}^{k,l}f  \nonumber
\\
&&+{}\lambda _{+}^{-1}\sum_{0\leq k+l<\infty }\alpha
_{+}^{k+l+1}\sum_{i=0}^{k}\sum_{j=0}^{l+1}\sum_{0\leq u\leq
i+j}(M^{+-})_{i,j,u}^{k,l+1}(\underline{x})(T_{5}^{0})_{u}^{k,l}f  \nonumber
\end{eqnarray}
where 
\begin{equation}
\alpha _{+}=-q^{2}\frac{\lambda ^{2}}{\lambda _{+}^{2}},\quad \beta
=q+\lambda _{+}.
\end{equation}
To get expressions with a more obvious structure we have introduced the
abbreviations 
\begin{eqnarray}
\hspace{-0.23in}(\tilde{T}^{3})_{j}^{i}f &=&\Big[ \left. (\tilde{O}%
^{3})_{i}f\right| _{x^{3}\rightarrow x^{0}+\tilde{x}^{3}}\Big] (q^{2j}%
\tilde{x}^{3}), \\[0.16in]
\hspace{-0.23in}(T_{1}^{-})_{j}^{i}f &=&\Big[ \left.
(O_{1}^{-})_{i}f\right| _{x^{3}\rightarrow x^{0}+\tilde{x}^{3}}\Big]
(q^{2(j+1)}\tilde{x}^{3}), \\
\hspace{-0.23in}(T_{2}^{-})_{j}^{i}f &=&\Big[ \left.
(O_{2}^{-})_{i}f\right| _{x^{3}\rightarrow x^{0}+\tilde{x}^{3}}\Big]
(q^{2j}\tilde{x}^{3}),  \nonumber \\[0.16in]
\hspace{-0.23in}(T_{1}^{0})_{j}^{i}f &=&\Big[ \left.
(O_{1}^{0})_{i}f\right| _{x^{3}\rightarrow y_{+}}\hspace{-0.04in}-q^{2}\frac{%
\lambda }{\lambda _{+}}\left. (O_{2}^{0})_{i}f\right| _{x^{3}\rightarrow
q^{2}y_{+}}\Big] (q^{2j}\tilde{x}^{3}), \\
\hspace{-0.23in}(T_{2}^{0})_{j}^{i}f &=&\Big[ \left.
(O_{3}^{0})_{i}f\right| _{x^{3}\rightarrow x^{0}+\tilde{x}^{3}}+\left.
(O_{4}^{0})_{i}f\right| _{x^{3}\rightarrow q^{2}y_{-}}\Big] (q^{2j}\tilde{x%
}^{3}),  \nonumber \\
\hspace{-0.23in}(T_{3}^{0})_{u}^{k,l}f &=&\Big[ \left.
(Q_{1}^{0})_{k,l}f\right| _{x^{3}\rightarrow x^{0}+\tilde{x}^{3}}\Big]
(q^{2u}\tilde{x}^{3}),  \nonumber \\
\hspace{-0.23in}(T_{4}^{0})_{u}^{k,l}f &=&\Big[ \left.
(Q_{2}^{0})_{k,l}f\right| _{x^{3}\rightarrow x^{0}+\tilde{x}^{3}}\Big]
(q^{2u}\tilde{x}^{3}),  \nonumber \\
\hspace{-0.23in}(T_{5}^{0})_{u}^{k,l}f &=&\Big[ \left.
(Q_{3}^{0})_{k,l}f\right| _{x^{3}\rightarrow x^{0}+\tilde{x}^{3}}\Big]
(q^{2u}\tilde{x}^{3}),  \nonumber \\[0.16in]
\hspace{-0.23in}(T_{1}^{+})_{j}^{i}f &=&\Big[ \left.
(O_{1}^{+})_{i}f\right| _{x^{3}\rightarrow q^{2}y_{-}}\Big] (q^{2j}\tilde{x%
}^{3}), \\
\hspace{-0.23in}(T_{2}^{+})_{j}^{i}f &=&\Big[ \left.
(O_{2}^{+})_{i}f\right| _{x^{3}\rightarrow y_{+}}\Big] (q^{2j}\tilde{x}%
^{3}),  \nonumber \\
\hspace{-0.23in}(T_{3}^{+})_{u}^{k,l}f &=&\Big[ \left.
(Q_{1}^{+})_{k,l}f\right| _{x^{3}\rightarrow x^{0}+\tilde{x}^{3}}\Big]
(q^{2u}\tilde{x}^{3}),  \nonumber \\
\hspace{-0.23in}(T_{4}^{+})_{u}^{k,l}f &=&\Big[ \left.
(Q_{2}^{+})_{k,l}f\right| _{x^{3}\rightarrow x^{0}+\tilde{x}^{3}}\Big]
(q^{2u}\tilde{x}^{3})  \nonumber
\end{eqnarray}
which, in turn, depend on the following operators 
\begin{eqnarray}
( \tilde{O}^{3}) _{k}f &=&(D_{2,q}^{3})^{k,k+1}f(q^{2}x^{+}), \\%
[0.16in]
(O_{1}^{-})_{k}f &=&x^{-}(D_{2,q}^{3})^{k+1,k+1}f(q^{2}x^{+}), \\
(O_{2}^{-})_{k}f &=&\tilde{x}%
^{3}D_{q^{2}}^{+}(D_{2,q}^{3})^{k,k+1}f(q^{2}x^{+}),  \nonumber \\[0.16in]
(O_{1}^{0})_{k}f &=&\tilde{D}_{q^{2}}^{3}(D_{1,q}^{3})^{k,k}f \\
&&-\,q^{3}\lambda _{+}^{-1}\lambda ^{2}x^{+}\tilde{x}^{3}D_{q^{2}}^{+}\tilde{%
D}_{q^{2}}^{3}(D_{1,q}^{3})^{k,k+1}f,  \nonumber \\
(O_{2}^{0})_{k}f &=&x^{-}(D_{1,q^{-1}}^{3})^{k,k+1}D_{q^{2}}^{-}f(q^{2}%
\tilde{x}^{3}),  \nonumber \\
(O_{3}^{0})_{k}f &=&qx^{+}D_{q^{2}}^{+}(D_{2,q}^{3})^{k,k+1}f,  \nonumber \\
(O_{4}^{0})_{k}f &=&\tilde{x}%
^{3}D_{q^{2}}^{+}(D_{1,q^{-1}}^{3})^{k,k}D_{q^{2}}^{-}f,  \nonumber \\%
[0.16in]
(Q_{1}^{0})_{k,l}f &=&(x^{0}+q^{-1}\lambda
x^{3})(D_{3,q}^{3})_{l,l+1}^{k+1,k}f \\
&&+\,q\lambda _{+}^{-1}\lambda (q+\lambda _{+})x^{+}\tilde{x}%
^{3}D_{q^{2}}^{+}(D_{3,q}^{3})_{l,l+1}^{k,k+1}f  \nonumber \\
&&-\,q^{3}\lambda _{+}^{-1}\lambda ^{2}x^{+}\tilde{x}^{3}(x^{0}+q^{-1}%
\lambda \tilde{x}^{3})D_{q^{2}}^{+}(D_{3,q}^{3})_{l,l+1}^{k+1,k+1}f, 
\nonumber \\
(Q_{2}^{0})_{k,l}f &=&(D_{3,q}^{3})_{l,l+1}^{k+1,k+1}f,  \nonumber \\
(Q_{3}^{0})_{k,l}f &=&q^{-1}(D_{3,q}^{3})_{l+1,l+1}^{k+1,k}f-q^{2}\lambda
_{+}^{-1}\lambda ^{2}x^{+}\tilde{x}%
^{3}D_{q^{2}}^{+}(D_{3,q}^{3})_{l+1,l+1}^{k+1,k+1}f,  \nonumber \\[0.16in]
(O_{1}^{+})_{k}f &=&(D_{1,q^{-1}}^{3})^{k,k}D_{q^{2}}^{-}f, \\
(O_{2}^{+})_{k}f &=&x^{+}\tilde{D}_{q^{2}}^{3}(D_{1,q}^{3})^{k,k+1}f, 
\nonumber \\[0.16in]
(Q_{1}^{+})_{k,l}f &=&(q+\lambda _{+})x^{+}(D_{3,q}^{3})_{l,l+1}^{k,k+1}f \\
&&-\,q^{2}\lambda x^{+}(x^{0}+q^{-1}\lambda
x^{3})(D_{3,q}^{3})_{l,l+1}^{k+1,k+1}f,  \nonumber \\
(Q_{2}^{+})_{k,l}f &=&x^{+}(D_{3,q}^{3})_{l+1,l+1}^{k+1,k+1}f.  \nonumber
\end{eqnarray}
And in the same way the representations of the Lorentz generators are
explicitely given by 
\begin{eqnarray}
&&\Lambda \triangleright f(x^{+},x^{0},\tilde{x}^{3},x^{-})= 
f(q^{-2}x^{+},q^{-2}x^{0},q^{-2}\tilde{x}^{3},q^{-2}x^{-}),
  \\[0.16in]
&&\tau ^{3}\triangleright f(x^{+},x^{0},\tilde{x}^{3},x^{-})= 
f(q^{-4}x^{+},q^{4}x^{-}),   \\[0.16in]
&&T^{+}\triangleright f(x^{+},x^{0},\tilde{x}^{3},x^{-})= \\
&&q^{-\frac{1}{2}}\lambda _{+}^{\frac{1}{2}}\left[ x^{0}D_{q^{2}}^{-}f+%
\tilde{x}^{3}D_{q^{4}}^{-}f+qx^{+}\tilde{D}_{q^{2}}^{3}f\right]
(q^{-2}x^{+}),  \nonumber \\[0.16in]
&&T^{-}\triangleright f(x^{+},x^{0},\tilde{x}^{3},x^{-})= \\
&&q^{\frac{1}{2}}\lambda _{+}^{\frac{1}{2}}\left[ x^{0}D_{q^{2}}^{+}f+%
\tilde{x}^{3}D_{q^{4}}^{+}f+qx^{-}\tilde{D}_{q^{2}}^{3}f\right]
(q^{-2}x^{+}),  \nonumber \\[0.16in]
&&T^{2}\triangleright f(x^{+},x^{3},\tilde{x}^{3},x^{-})= \\
&&\lambda _{+}^{-\frac{1}{2}}\sum_{k=0}^{\infty }\alpha _{0}^{k}\sum_{0\leq
i+j\leq k}\Big\{ q^{\frac{1}{2}}(M^{-})_{i,j}^{k}(\underline{x}%
)(T_{1}^{T})_{j}^{i}f+q^{-\frac{1}{2}}(M^{+})_{i,j}^{k}(\underline{x}%
)(T_{2}^{T})_{j}^{i}f\Big\},   \nonumber \\[0.16in]
&&S^{1}\triangleright f(x^{+},x^{3},\tilde{x}^{3},x^{-})= \\
&&-q\lambda _{+}^{-\frac{1}{2}}\sum_{k=0}^{\infty }\alpha
_{0}^{k}\sum_{0\leq i+j\leq k}\Big\{ q^{\frac{1}{2}}(M^{-})_{i,j}^{k}(%
\underline{x})(T_{1}^{S})_{j}^{i}f+q^{-\frac{1}{2}}(M^{+})_{i,j}^{k}(%
\underline{x})(T_{2}^{S})_{j}^{i}f\Big\},   \nonumber \\[0.16in]
&&\tau ^{1}\triangleright f(x^{+},x^{3},\tilde{x}^{3},x^{-})= \\
&&\sum_{k=0}^{\infty }\alpha _{0}^{k}\sum_{0\leq i+j\leq k}\Big\{
(M^{+})_{i,j}^{k}(\underline{x})(T_{2}^{\tau })_{j}^{i}f-\frac{\lambda ^{2}}{%
\lambda _{+}}(M^{-})_{i,j}^{k}(\underline{x})(T_{1}^{\tau
})_{j}^{i}f\Big\},   \nonumber \\[0.16in]
&&\sigma ^{2}\triangleright f(x^{+},x^{3},\tilde{x}^{3},x^{-})= \\
&&\sum_{k=0}^{\infty }\alpha _{0}^{k}\sum_{0\leq i+j\leq\
k}(M^{-})_{i,j}^{k}(\underline{x})(T^{\sigma })_{j}^{i}f  \nonumber
\end{eqnarray}
where $\alpha _{0}=-(\lambda /\lambda _{+})^{2}.$ For the purpose of
abbreviation we have again set 
\begin{eqnarray}
(T_{1}^{T})^k_j\triangleright f &=&\Big[ \left. (O_{1}^{T})_k f\right|
_{x^{3}\rightarrow y_{-}}\Big] (q^{2j}\tilde{x}^{3}), \\
(T_{2}^{T})^k_j\triangleright f &=&\Big[ \left. (O_{2}^{T})_k f\right|
_{x^{3}\rightarrow y_{+}}\Big] (q^{2j}\tilde{x}^{3}),  \nonumber \\[0.16in]
(T_{1}^{S})^k_j\triangleright f &=&\Big[ \left. (O_{1}^{S})_k f\right|
_{x^{3}\rightarrow y_{-}}\Big] (q^{2j}\tilde{x}^{3}), \\
(T_{2}^{S})^k_j\triangleright f &=&\Big[ \left. (O_{2}^{S})_k f\right|
_{x^{3}\rightarrow y_{+}}\Big] (q^{2j}\tilde{x}^{3}),  \nonumber \\[0.16in]
(T_{1}^{\tau })^k_j\triangleright f &=&\Big[ \left. (O_{1}^{\tau })_k f\right|
_{x^{3}\rightarrow y_{-}}\Big] (q^{2j}\tilde{x}^{3}), \\
(T_{2}^{\tau })^k_j\triangleright f &=&\Big[ \left. (O_{2}^{\tau })_k f\right|
_{x^{3}\rightarrow y_{+}}-q\frac{\lambda ^{2}}{\lambda _{+}}\left.
(O_{3}^{\tau })_k f\right| _{x^{3}\rightarrow y_{+}}\Big] (q^{2j}\tilde{x}%
^{3}),  \nonumber \\[0.16in]
(T^{\sigma })^k_j\triangleright f &=&\Big[ \left. (O^{\sigma })f\right|
_{x^{3}\rightarrow y_{-}}\Big] (q^{2j}\tilde{x}^{3})
\end{eqnarray}
where 
\begin{eqnarray}
(O_{1}^{T})_{k}f &=&q^{-1}\tilde{x}%
^{3}(D_{1,q^{-1}}^{3})^{k,k}D_{q^{2}}^{-}f(qx^{+},q^{-1}\tilde{x}%
^{3},q^{-1}x^{-}), \\
(O_{2}^{T})_{k}f &=&x^{+}(D_{1,q}^{3})^{k,k+1}f(qx^{+},q\tilde{x}%
^{3},q^{-1}x^{-}),  \nonumber \\[0.16in]
(O_{1}^{S})_{k}f &=&q^{-1}\tilde{x}%
^{3}(D_{1,q^{-1}}^{3})^{k,k}D_{q^{2}}^{+}f(q^{-1}x^{+},q^{-1}\tilde{x}%
^{3},qx^{-}), \\
(O_{2}^{S})_{k}f &=&x^{-}(D_{1,q^{-1}}^{3})^{k,k+1}f(q^{-1}x^{+},q\tilde{x}%
^{3},qx^{-}),  \nonumber \\[0.16in]
(O_{1}^{\tau })_{k}f &=&(\tilde{x}%
^{3})^{2}D_{q^{2}}^{+}(D_{1,q^{-1}}^{3})^{k,k}D_{q^{2}}^{-}f(qx^{+},q^{-1}%
\tilde{x}^{3},q^{-1}x^{-}), \\
(O_{2}^{\tau })_{k}f &=&(D_{1,q}^{3})^{k,k}f(qx^{+},q\tilde{x}%
^{3},q^{-1}x^{-})  \nonumber \\
&&-\;{}q\lambda _{+}^{-1}\lambda ^{2}x^{+}\tilde{x}%
^{3}D_{q^{2}}^{+}(D_{1,q}^{3})^{k,k+1}f(qx^{+},q\tilde{x}^{3},q^{-1}x^{-}), 
\nonumber \\
(O_{3}^{\tau })_{k}f &=&q\tilde{x}%
^{3}x^{-}(D_{1,q^{-1}}^{3})^{k,k+1}D_{q^{2}}^{-}f(qx^{+},q\tilde{x}%
^{3},q^{-1}x^{-}),  \nonumber \\[0.16in]
(O^{\sigma })_{k}f &=&(D_{1,q^{-1}}^{3})^{k,k}f(q^{-1}x^{+},q^{-1}\tilde{x}%
^{3},qx^{-}).
\end{eqnarray}

Now we deal on with representations for the partial derivatives $\partial
^{\mu }$, $\mu =0,\pm ,\tilde{3}$. These can directly be obtained from the
representations of the conjugated ones, if we apply the following
transformation 
\begin{eqnarray}
&&\partial ^{\pm }\tilde{\triangleright}f\stackrel{{\QATOP{\pm }{q}}{\QATOP{%
\leftrightarrow }{\leftrightarrow }}{\QATOP{\mp }{1/q}}}{\longleftrightarrow 
}\hat{\partial}^{\mp }\triangleright f,  \label{transrulrl} \\
&&\partial ^{3}\tilde{\triangleright}f\stackrel{{\QATOP{\pm }{q}}{\QATOP{%
\leftrightarrow }{\leftrightarrow }}{\QATOP{\mp }{1/q}}}{\longleftrightarrow 
}\hat{\partial}^{3}\triangleright f  \nonumber
\end{eqnarray}
which means concretely, that the substitutions 
\begin{equation}
D_{q^{a}}^{\pm }\rightarrow D_{q^{-a}}^{\mp },\quad \hat{n}_{\pm
}\rightarrow -\hat{n}_{\mp },\quad q^{\pm 1}\rightarrow q^{\mp 1}
\end{equation}
interchange the different representations. It is very important to
notice that the representations on the left hand side of (\ref{transrulrl})
have to refer to a different normal ordering given by
$X^{-}X^{3}\tilde{X}^{3}X^{+}.$ However, by the identities 
\begin{eqnarray}
\partial ^{\mu }\triangleright (\hat{U}^{-1}f) &=&\hat{U}^{-1}(\partial
^{\mu }\tilde{\triangleright}f), \\
\hat{\partial}^{\mu }\tilde{\triangleright}(\hat{U}f) &=&\hat{U}(\hat{%
\partial}^{\mu }\triangleright f)  \nonumber
\end{eqnarray}
with 
\begin{eqnarray}
&&\hat{U}^{-1}f=  \label{UmMink1} \\
&&\sum_{i=0}^{\infty }\left( \frac{\lambda }{\lambda _{+}}\right)
^{i}\sum_{k+j=i}\frac{(R_{q})_{k,j}(\underline{x})}{%
[[k]]_{q^{2}}![[j]]_{q^{2}}!}q^{2\hat{n}_{+}\hat{n}_{-}+(\hat{n}_{+}+\hat{n}%
_{-})(2\hat{n}_{3}+i)+2\hat{n}_{3}i}  \nonumber \\[0.16in]
&&\cdot \left( (D_{q^{2}}^{+})^{i}(D_{q^{2}}^{-})^{i}f\right)
(x^{0},q^{j-k}x^{+},\tilde{x}^{3},q^{j-k}x^{-}),  \nonumber \\[0.16in]
&&\hat{U}f=  \label{UmMin2} \\
&&\sum_{i=0}^{\infty }\left( -\frac{\lambda }{\lambda _{+}}\right)
^{i}\sum_{k+j=i}\frac{(R_{q^{-1}})_{k,j}(\underline{x})}{%
[[k]]_{q^{-2}}![[j]]_{q^{-2}}!}q^{-2\hat{n}_{+}\hat{n}_{-}-(\hat{n}_{+}+\hat{%
n}_{-})(2\hat{n}_{3}+i)-2\hat{n}_{3}i}  \nonumber \\
&&\cdot \left( (D_{q^{-2}}^{+})^{i}(D_{q^{-2}}^{-})^{i}f\right)
(x^{0},q^{k-j}x^{+},\tilde{x}^{3},q^{k-j}x^{-})  \nonumber
\end{eqnarray}
and {\cite{WW01}}
\begin{equation}
(R_q)_{k,j}(\underline{x})=(-q)^k(q^j(\tilde{x}^3)^2)^j
\sum_{p=0}^{k}(S_q)_{k,p}(x^0,\tilde{x}^3)(q^{4j}\lambda_+x^+x^-)^p 
\end{equation}
the representations for ordering $X^{-}X^{3}\tilde{X
}^{3}X^{+}$ can be transformed into those for reversed ordering 
$X^{+}X^{3}\tilde{X}^{3}X^{-}$.
The reason for the existence of (\ref{transrulrl}) should be clear
from the fact that the substitutions 
\begin{equation}
\partial ^{\pm }\rightarrow \hat{\partial}^{\mp },\quad \tilde{\partial}%
^{3}\rightarrow \hat{\tilde{\partial}^{3}},\quad \partial ^{0}\rightarrow 
\hat{\partial}^{0}\quad X^{\pm }\rightarrow X^{\mp },\quad q\rightarrow
q^{-1}
\end{equation}
interchange the relations (\ref{AblKoordMinAnf} - \ref{AblKoordMin End}) and (%
\ref{D3Min} - \ref{DEnde}).

Finally, from the conjugation properties \cite{LWW97} 
\begin{equation}
\overline{X^{0}}=X^{0},\quad \overline{\tilde{X}^{3}}=\tilde{X}^{3},\quad 
\overline{X^{+}}=-qX^{-},\quad \overline{X^{-}}=-q^{-1}X^{+},
\end{equation}
\begin{equation}
\overline{\partial ^{0}}=-q^{4}\hat{\partial}^{0},\quad \overline{\partial
^{3}}=-q^{4}\hat{\tilde{\partial}^{3}},\quad \overline{\partial ^{+}}=+q^{5}%
\hat{\partial}^{-},\quad \overline{\partial ^{-}}=+q^{3}\hat{\partial}^{+},
\end{equation}
\begin{equation}
\overline{T^{+}}=q^{-2}T^{-},\quad \overline{T^{-}}=q^{2}T^{+}
\end{equation}
we can derive the following
rules for transforming right and left representations into each other 
\begin{eqnarray}
&f\triangleleft \partial ^{0}&\stackrel{+\leftrightarrow -}{%
\longleftrightarrow }-q^{4}\hat{\partial}^{0}\triangleright f, \\
&f\triangleleft \tilde{\partial}^{3}&\stackrel{+\leftrightarrow -}{%
\longleftrightarrow }-q^{4}\hat{\tilde{\partial}^{3}}\triangleright f, 
\nonumber \\
&f\triangleleft \partial ^{+}&\stackrel{+\leftrightarrow -}{%
\longleftrightarrow }-q^{4}\hat{\partial}^{-}\triangleright f,  \nonumber \\
&f\triangleleft \partial ^{-}&\stackrel{+\leftrightarrow -}{%
\longleftrightarrow }-q^{4}\hat{\partial}^{+}\triangleright f,  \nonumber \\%
[0.16in]
&f\triangleleft \hat{\partial}^{0}&\stackrel{+\leftrightarrow -}{%
\longleftrightarrow }-q^{-4}\partial ^{0}\triangleright f, \\
&f\triangleleft \hat{\tilde{\partial}^{3}}&\stackrel{+\leftrightarrow -}{%
\longleftrightarrow }-q^{-4}\tilde{\partial}^{3}\triangleright f,  \nonumber
\\
&f\triangleleft \hat{\partial}^{+}&\stackrel{+\leftrightarrow -}{%
\longleftrightarrow }-q^{-4}\partial ^{-}\triangleright f,  \nonumber \\
&f\triangleleft \hat{\partial}^{-}&\stackrel{+\leftrightarrow -}{%
\longleftrightarrow }-q^{-4}\partial ^{+}\triangleright f,  \nonumber \\%
[0.16in]
&f\triangleleft T^{+}&\stackrel{+\leftrightarrow -}{\longleftrightarrow }%
-q^{-3}T^{-}\triangleright f, \\
&f\triangleleft T^{-}&\stackrel{+\leftrightarrow -}{\longleftrightarrow }%
-q^{3}T^{+}\triangleright f \nonumber
\end{eqnarray}
where the symbol $\stackrel{+\leftrightarrow -}{\longleftrightarrow }$
 has the same meaning as in section \ref{sec2}.
Once again the simplest way to determine right representations for the
remaining generators is described by the identity 
\begin{equation}
f\triangleleft h=S^{-1}(h)\triangleright f.
\end{equation}
With the Hopf structure of the Lorentz generators at hand \cite{Blo01} we
end up with the expressions 
\begin{eqnarray}
f\triangleleft T^{2} &=&-(\tau ^{3})^{\frac{1}{2}}T^{2}\triangleright f, \\
f\triangleleft S^{1} &=&-q^{2}(\tau ^{3})^{-\frac{1}{2}}S^{1}\triangleright
f,  \nonumber \\[0.16in]
f\triangleleft \tau ^{1} &=&\sigma ^{2}\triangleright f, \\
f\triangleleft \sigma ^{2} &=&\tau ^{1}\triangleright f,  \nonumber \\%
[0.16in]
f\triangleleft \tau ^{3} &=&(\tau ^{3})^{-1}\triangleright f, \\
f\triangleleft \Lambda  &=&\Lambda ^{-1}\triangleright f.  \nonumber
\end{eqnarray}

\section{Remarks}

Let us end with a few comments on our representations. First of all, from a
physical point of view partial derivatives are objects generating
translations in time or space. According to 
\begin{equation}
\partial ^{A}f=(\partial ^{A}f)_{0}+\sum_{i>0}\lambda ^{i}(\partial
^{A}f)_{i}
\end{equation}
their representations can be divided up into one part reducing to ordinary
derivatives in the undeformed limit (q=1) \ and a second part of correction
terms disappearing in that case. The existence of the correction terms can
be well understood, if one assumes that non-commutativity results from a
coupling of the different directions in space. Thus a flow of momentum in
only one direction is in general not possible and the corrections should be
responsible for this feature. So far we can sum up that the situation in
non-commutative spaces seems to be similar to that of solids.
In fact, if such a solid state has to undergo a deformation in some
direction, the other directions will also be influenced due to their coupling.

\appendix 

\section{Notation\label{AppA}}
\begin{enumerate}
\item  The \textit{q-number} is defined by \cite{KS97} 
\begin{equation}
\left[ \left[ c\right] \right] _{q^{a}}\equiv \frac{1-q^{ac}}{1-q^{a}}%
,\qquad a,c\in \mathbb{C}.
\end{equation}
\item  For $m\in \mathbb{N}$, we can introduce the \textit{q-factorial }by
setting 
\begin{equation}
\left[ \left[ m\right] \right] _{q^{a}}!\equiv \left[ \left[ 1\right]
\right] _{q^{a}}\left[ \left[ 2\right] \right] _{q^{a}}\ldots \left[ \left[
m\right] \right] _{q^{a}},\qquad \left[ \left[ 0\right] \right]
_{q^{a}}!\equiv 1.
\end{equation}
\item  There is also a q-analogue of the usual binomial coefficients, the
so-called \textit{q-binomial coefficients} defined by the formula 
\begin{equation}
\QATOPD {\alpha }{m}_{q^{a}}\equiv \frac{\left[ \left[ \alpha \right]
\right] _{q^{a}}\left[ \left[ \alpha -1\right] \right] _{q^{a}}\ldots \left[
\left[ \alpha -m+1\right] \right] _{q^{a}}}{\left[ \left[ m\right] \right]
_{q^{a}}!}
\end{equation}
where $\alpha \in \mathbb{C},$ $m\in \mathbb{N}$.
\item  \noindent Commutative co-ordinates are usually denoted by small 
letters (e.g. $x^{+},$ $x^{-},$ etc.), non-commutative
co-ordinates in capital (e.g. $X^{+},$ $X^{-},$ etc.).
\item  \noindent Note that in functions only such arguments are explicitly
displayed which are effected by a scaling. For example, we write
\begin{equation}
f(q^{2}x^{+})\qquad \text{instead of}\qquad f(q^{2}x^{+},x^{3},x^{-}).
\end{equation}
\item  \noindent Arguments in parentheses shall refer to the first object on
their left. For example, we have 
\begin{equation}
D_{q^{2}}^{+}f(q^{2}x^{+})=D_{q^{2}}^{+}(f(q^{2}x^{+}))
\end{equation}
or 
\begin{equation}
D_{q^{2}}^{+}\left[ D_{q^{2}}^{+}f+D_{q^{2}}^{-}f\right]
(q^{2}x^{+})=D_{q^{2}}^{+}\left( \left[ D_{q^{2}}^{+}f+D_{q^{2}}^{-}f\right]
(q^{2}x^{+})\right) .
\end{equation}
However, the symbol $\mid _{x^{\prime }\rightarrow x}$ applies to the whole
expression on its left side reaching up to the next opening bracket or $\pm $
sign.
\item  \noindent The \textit{Jackson derivative} referring to the coordinate 
$x^{A}$ is defined by 
\begin{equation}
D_{q^{a}}^{A}f\equiv \frac{f(x^{A})-f(q^{a}x^{A})}{(1-q^{a})x^{A}}
\end{equation}
where f may depend on other coordinates as well. Higher Jackson
derivatives are obtained by applying the above operator $D_{q^{a}}^{A}$
several times: 
\begin{equation}
(D_{q^{a}}^{A})^{i}f\equiv \underbrace{D_{q^{a}}^{A}D_{q^{a}}^{A}\ldots
D_{q^{a}}^{A}}_{i\text{ times}}f.
\end{equation}
\item  Additionally, we need operators of the following form 
\begin{equation}
\hat{n}^{A}\equiv x^{A}\frac{\partial }{\partial x^{A}}.
\end{equation}
\end{enumerate}

\section{New Jackson Derivatives \label{AppB}}

In the calculations of chapter \ref{chap4} we have introduced the following
quantities 
\begin{equation}
(K_{n})_{a}^{(k)}\equiv \sum_{j_{1}=0}^{n-k}\sum_{j_{2}=0}^{n-k-j_{1}}\ldots
\sum_{j_{k}=0}^{n-k-j_{1}-\ldots -j_{k-1}}a^{j_{1}+j_{2}+\ldots
+j_{k}},\quad n\geq k\geq 1.
\end{equation}
Additionally, we have defined an operation $\circ $ by
\begin{equation}
(K_{n})_{a}^{(k)}\circ (K_{n-k})_{b}^{(l)}\equiv
\sum_{j_{1}=0}^{n-k-l}\ldots \sum_{j_{k+l}=0}^{n-k-l-j_{1}-\ldots
-j_{k+l-1}}a^{j_{1}+\ldots +j_{k}}b^{j_{k+1}+\ldots +j_{k+l}}
\end{equation}
leading to new expressions denoted by
\begin{equation}
(K_{n})_{a_{1},\ldots ,a_{l}}^{(k_{1},\ldots ,k_{l})}\equiv
(K_{n})_{a_{1}}^{(k_{1})}\circ (K_{n-k_{1}})_{a_{2}}^{(k_{2})}\circ \ldots
\circ (K_{n-k_{1}-\ldots -k_{l}})_{a_{l}}^{(k_{l})}.  \label{VarNeu}
\end{equation}
Furthermore these new quantities show a number of simple properties, for
instance,
\begin{enumerate}
\item  \begin{equation}
(K_{n})_{1}^{(k)}=\binom{n}{k}.
\end{equation}
\item  
\begin{equation}
(K_{n})_{a,a}^{(k,l)}=(K_{n})_{a}^{(k+l)}.
\end{equation}
\item  
\begin{equation}
(K_{n})_{a_{1},\ldots ,a_{l}}^{(k_{1},\ldots ,k_{l})}=(K_{n})_{a_{\pi
(1)},\ldots ,a_{\pi (l)}}^{(k_{\pi (1)},\ldots ,k_{\pi (l)})},
\end{equation}
where $\pi $ is any permutation of the $\{1,\ldots ,l\}$.
\end{enumerate}
Due to these properties the quantities of (\ref{VarNeu}) can be divided up
into three different sets:
\begin{enumerate}
\item  
\begin{equation}
(K_{n})_{a_{1},\ldots ,a_{l}}^{(k_{1},\ldots ,k_{l})},\quad a_{i}\neq 1\quad 
\text{for all }i\in \{1,\ldots ,l\},  \label{ersteKat}
\end{equation}
\item  
\begin{equation}
(K_{n})_{a_{1},\ldots ,a_{l-1},1}^{(k_{1},\ldots ,k_{l})},\quad a_{i}\neq
1\quad \text{for all }i\in \{1,\ldots ,l-1\},  \label{zweiteKat}
\end{equation}
\item  
\begin{equation}
(K_{n})_{1}^{(k)}.
\end{equation}
\end{enumerate}
It is now our aim to show that each $(K_{n})_{a_{1},\ldots
,a_{l}}^{(k_{1},\ldots ,k_{l})}$ can be expressed in terms of the binomials $%
(K_{n})_{1}^{(k)}.$ Towards this end we need the following additional rules
\begin{enumerate}
\item  
\begin{equation}
(K_{n})_{a}^{(k)}=(1-a)^{-k}-%
\sum_{m=0}^{k-1}a^{n-m}(1-a)^{m-k}(K_{n})_{1}^{(m)},
\end{equation}
\item  
\begin{eqnarray}
(K_{n})_{a,1}^{(k,l)} &=&\sum_{m=0}^{l}(-1)^{l-m}\binom{k+l-1-m}{k-1}
\nonumber\\
&&\hspace{0.5in}\cdot (1-a)^{m-k-l}(K_{n})_{1}^{(m)} \\
&&+(-1)^{l+1}\sum_{m=0}^{k-1}a^{n-m}\binom{k+l-1-m}{l} \nonumber \\
&&\hspace{0.5in}\cdot (1-a)^{m-k-l}(K_{n})_{1}^{(m)},  \nonumber
\end{eqnarray}
\item  
\begin{eqnarray}
&&(K_{n})_{a_{1},\ldots ,a_{l}}^{(k_{1},\ldots ,k_{l})}\circ \Big[ (
b) ^{n-k_{1}-\ldots -k_{l}}(K_{n-k_{1}-\ldots -k_{l}})_{c_{1},\ldots
,c_{p}}^{(q_{1},\ldots ,q_{p})}\Big]=  \\
&&b^{n-k_{1}-\ldots -k_{l}}\Big[ (K_{n})_{\frac{a_{1}}{b},\ldots ,\frac{%
a_{l}}{b}}^{(k_{1},\ldots ,k_{l})}\circ (K_{n-k_{1}-\ldots
-k_{l}})_{c_{1},\ldots ,c_{p}}^{(q_{1},\ldots ,q_{p})}\Big]   \nonumber
\end{eqnarray}
\end{enumerate}
which can be verified quite elementarily. From these rules we find a recursion
relation for which we have
\begin{eqnarray}
&&(K_{n})_{a_{1},\ldots ,a_{l}}^{(k_{1},\ldots ,k_{l})}= \\
&&(K_{n})_{a_{1},\ldots ,a_{l-1}}^{(k_{1},\ldots ,k_{l-1})}\circ
(K_{n-k_{1}-\ldots -k_{l-1}})_{a_{l}}^{(k_{l})}=  \nonumber \\
&&-\sum_{m=0}^{k_{l}-1}a_{l}^{n-k_{1}-\ldots
-k_{l-1}-m}(1-a_{l})^{m-k_{l}}(K_{n})_{\frac{a_{1}}{a_{l}},\ldots ,\frac{%
a_{l-1}}{a_{l}},1}^{(k_{1},\ldots ,k_{l-1},m)}  \nonumber \\
&&+(1-a_{l})^{-k_{l}}(K_{n})_{a_{1},\ldots ,a_{l-1}}^{(k_{1},\ldots
,k_{l-1})}.  \nonumber
\end{eqnarray}
Using this relation repeatedly the quantities of the first set can now be
reduced to those of the second one, as one gets
\begin{eqnarray}
&&(K_{n})_{a_{1},\ldots ,a_{l}}^{(k_{1},\ldots ,k_{l})}=  \label{RekuKom1} \\
&&-\sum_{i=1}^{l}\Big( \prod_{j=i+1}^{l}(1-a_{j})^{-k_{j}}\Big)  
\nonumber \\
&&\cdot \sum_{m=0}^{k_{i}-1}a_{i}^{n-k_{1}-\ldots
-k_{i-1}-m}(1-a_{i})^{m-k_{i}}(K_{n})_{\frac{a_{1}}{a_{i}},\ldots ,\frac{%
a_{i-1}}{a_{i}},1}^{(k_{1},\ldots ,k_{i-1},m)}.  \nonumber
\end{eqnarray}
In the same way the quantities of the second set can, in turn, be reduced to
those of the last one by applying another recursion relation given by
\begin{eqnarray}
&&(K_{n})_{a_{1},\ldots ,a_{l-1},1}^{(k_{1},\ldots ,k_{l})}=
\label{RekurKom2} \\
&&(K_{n})_{a_{1},\ldots ,a_{l-2}}^{(k_{1},\ldots ,k_{l-2})}\circ
(K_{n-k_{1}-\ldots -k_{l-2}})_{a_{l-1},1}^{(k_{l-1},k_{l})}=  \nonumber \\
&&\sum_{m=0}^{k_{l}}(-1)^{k_{l}-m}\binom{k_{l}+k_{l-1}-m-1}{k_{l-1}-1} 
\nonumber \\
&&\hspace{0.3in}\cdot\, (1-a_{l-1})^{m-k_{l}-k_{l-1}}(K_{n})_{a_{1},\ldots
,a_{l-2},1}^{(k_{1},\ldots ,k_{l-2},m)}  \nonumber \\
&&-(-1)^{k_{l}}\sum_{m=0}^{k_{l-1}-1}\binom{k_{l}+k_{l-1}-m-1}{k_{l}} 
\nonumber \\
&&\hspace{0.3in}\cdot\, a_{l-1}^{n-k_{1}-\ldots
-k_{l-2}-m}(1-a_{l-1})^{m-k_{l}-k_{l-1}}(K_{n})_{\frac{a_{1}}{a_{l-1}}%
,\ldots ,\frac{a_{l-2}}{a_{l-1}},1}^{(k_{1},\ldots ,k_{l-2},m)}.  \nonumber
\end{eqnarray}
If we introduce operators acting on powers $x^{n}$ by
\begin{equation}
D_{a_{1},\ldots ,a_{l}}^{(k_{1},\ldots ,k_{l})}x^{n}=
(K_{n})_{a_{1},\ldots ,a_{l}}^{(k_{1},\ldots ,k_{l})}x^{n-k_{1}-\ldots
-k_{l}}{,\quad \text{if }n\le k_{1}+\ldots +k_{l},}
\end{equation}
the relations (\ref{RekuKom1}) and (\ref{RekurKom2}) correspond to the
identities
\begin{eqnarray}
&&D_{a_{1},\ldots ,a_{l}}^{(k_{1},\ldots ,k_{l})}f(x)= \\
&&-\sum_{i=1}^{l}\Big( \prod_{j=i+1}^{l}(x-a_{j}x)^{-k_{j}}\Big)  
\nonumber \\
&&\cdot \sum_{m=0}^{k_{i}-1}(x-a_{i}x)^{m-k_{i}}\Big( D_{\frac{a_{1}}{a_{i}}%
,\ldots ,\frac{a_{i-1}}{a_{i}},1}^{(k_{1},\ldots ,k_{i-1},m)}f\Big)
(a_{i}x)  \nonumber
\end{eqnarray}
where $a_{i}\neq 1$ for all $i\in \{1,\ldots ,l\},$ and
\begin{eqnarray}
&&D_{a_{1},\ldots ,a_{l-1},1}^{(k_{1},\ldots ,k_{l})}f(x)= \\
&&\sum_{m=0}^{k_{l}}(-1)^{k_{l}-m}\binom{k_{l}+k_{l-1}-m-1}{k_{l-1}-1} 
\nonumber \\
&&\hspace{0.3in}\cdot\, (x-a_{l-1}x)^{m-k_{l}-k_{l-1}}D_{a_{1},\ldots
,a_{l-2},1}^{(k_{1},\ldots ,k_{l-2},m)}f(x)  \nonumber \\
&&-(-1)^{k_{l}}\sum_{m=0}^{k_{l-1}-1}\binom{k_{l}+k_{l-1}-m-1}{k_{l}} 
\nonumber \\
&&\hspace{0.3in}\cdot\, (x-a_{l-1}x)^{m-k_{l}-k_{l-1}}\Big( D_{\frac{a_{1}}{%
a_{l-1}},\ldots ,\frac{a_{l-2}}{a_{l-1}},1}^{(k_{1},\ldots
,k_{l-2},m)}f\Big) (a_{l-1}x)  \nonumber
\end{eqnarray}
where $a_{i}\neq 1$ for all $i\in \{1,\ldots ,l-1\}.$ With these formulae at
hand one readily checks that the derivative operators $D_{a_{1},\ldots
,a_{l}}^{(k_{1},\ldots ,k_{l})}$ can always be expressed in terms of the
simple operators 
\begin{equation}
D_{1}^{(k)}f(x)=\frac{1}{k!}\frac{\partial ^{k}}{\partial x^{k}}f(x).
\end{equation}

\noindent \textbf{Acknowledgement}\newline
First of all we want to express our gratitude to Julius Wess for his
efforts, suggestions and discussions. And we would like to thank Michael
Wohlgenannt, Fabian Bachmeier, Christian Blohmann, and Marcus Gaul for
useful discussions and their steady support.

\end{document}